\documentclass[twocolumn,showpacs,prc,superscriptaddress,amsmath,amssymb,preprintnumbers]{revtex4-1}

\newcommand{ \be }{\begin{eqnarray}}
\newcommand{ \ee }{\end{eqnarray}}
\newcommand{ \la }{\langle}
\newcommand{ \ra }{\rangle}

\newcommand{ \ds }{\displaystyle}

\newcommand{ \mean }[1]{\la #1 \ra}   
\newcommand{ \dmean }[1]{\la\la #1 \ra\ra}   
\newcommand{ \sump }{\sideset{}{'}\sum}

\newcommand{ \psirp }{\Psi_{RP}}

\usepackage{dcolumn} 
\usepackage{bm} 

\usepackage{graphicx}
\usepackage{color}
\definecolor{dgreen}{cmyk}{1.,0.,1.,0.4}        
\definecolor{orange}{cmyk}{0.,0.353,1.,0.}    

\usepackage[pdftex,bookmarks]{hyperref}

\begin{document}
%

\title{
Flow analysis with cumulants: direct calculations
}
\author{Ante Bilandzic}
\affiliation{Nikhef, Science Park 105, 1098 XG Amsterdam, 
The Netherlands}
\affiliation{Utrecht University, P.O. Box 80000, 3508 TA Utrecht, 
The Netherlands}
\author{Raimond Snellings}
\affiliation{Utrecht University, P.O. Box 80000, 3508 TA Utrecht, 
The Netherlands}
\author{Sergei Voloshin} 
\affiliation{Wayne State University,  
666 W. Hancock Street,  Detroit,  MI 48201, USA }
\date{\today}

\begin{abstract}
Anisotropic flow measurements in heavy-ion collisions provide
important information on the properties of hot and dense matter. 
These measurements are based on analysis of azimuthal correlations
and might be biased by contributions from 
correlations that are not related to the initial geometry, 
so called non-flow.
To improve anisotropic flow measurements advanced methods based 
on multi-particle correlations (cumulants) have been developed to suppress 
non-flow contribution.  
These multi-particle correlations can be calculated by looping over
all possible multiplets, however this quickly becomes prohibitively 
CPU intensive.
Therefore, the most used technique for cumulant calculations is 
based on generating functions. 
This method involves approximations, and has its own biases,
 which complicates the interpretation of the results. 
In this paper we present a new exact method for direct calculations 
of multi-particle cumulants using moments of the flow vectors.
\end{abstract}


\maketitle

\section{Introduction}

\begin{figure}[thb]
 \begin{center}
   \includegraphics[width=0.3\textwidth]   {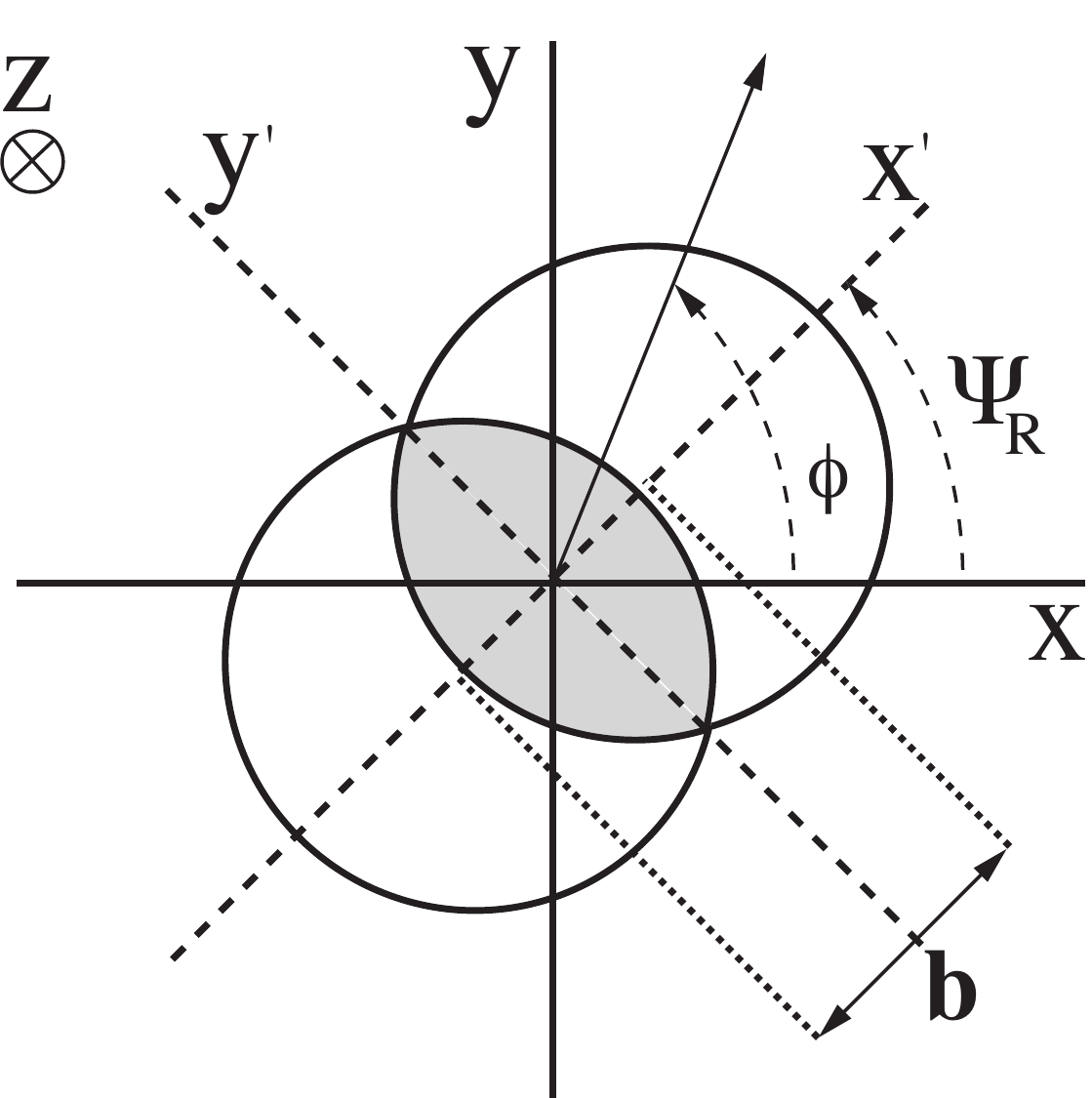}
   \caption{
     Schematic view of a non-central nucleus-nucleus collision in the
     transverse    plane.  
    }
    \label{geometry}
 \end{center}
\end{figure}

Anisotropic flow is a response of the system created in 
a heavy-ion collision to the anisotropies in the initial geometry. 
Thus, anisotropic flow is very sensitive to the properties 
of the system at an early time of its evolution.
The sizable azimuthal momentum-space anisotropy 
observed at RHIC energies (for a review, 
see~\cite{Voloshin:2008dg,Sorensen:2009cz})
is the main evidence for the nearly perfect liquid 
behavior~\cite{Teaney:2009qa,Heinz:2009xj} of the created matter. 
Quantitatively, anisotropic flow is characterized by 
coefficients in the Fourier expansion of the azimuthal dependence of the
invariant yield of particles relative to the reaction 
plane~\cite{Voloshin:1994mz,Poskanzer:1998yz}:
\begin{equation}
E\frac{d^3 N}{d^3 p} = \frac{1}{2\pi} \frac{d^2 N}{p_t dp_t
  dy}\! \left(1\!+\!\sum_{n=1}^{\infty}2v_n \cos\left(n\!\left(
\phi\!-\!\Psi_R\right) \right) \!\right). 
\label{eqFourier}
\end{equation}
Here $E$ is the energy of particle, 
$p_t$ is the transverse momentum, $\phi$ is 
its azimuthal angle, $y$ is the rapidity, and $\Psi_R$ the reaction
plane angle (see Fig~\ref{geometry}).  
The first coefficient, $v_1$, 
is usually called {\it directed flow}, and the second coefficient, 
$v_2$, is called {\it elliptic flow}. 
In general the $v_n = \mean{\cos[ n (\phi-\psirp)]}$ 
coefficients are $p_t$ and $y$ dependent -- in this context we refer 
to them as {\it differential  flow}. 
The {\it integrated} flow is defined as a weighted average with
the invariant distribution used as a weight:
\begin{equation}
v_n \equiv \frac{\ds \int_{0}^{\infty}
v_n(p_t) \frac{\ds dN}{\ds dp_t}dp_t}
{\ds \int_{0}^{\infty}\frac{\ds dN}{\ds dp_t}dp_t}\,.
\label{integratedFlow}
\end{equation}

Since the reaction plane $\Psi_R$ is not known experimentally, 
the anisotropic flow is estimated using azimuthal correlations between the
observed particles. 
For example, using 2-particle azimuthal correlations:
\begin{eqnarray}
\mean{\cos(n(\phi_1-\phi_2))}=
\mean{e^{in(\phi_{1} - \phi_{2})}} = \mean{v_{n}^{2}} + \delta_n,
\label{twoParticleFlowEstimate}
\end{eqnarray}
where the first term, $\mean{v_{n}^{2}}$, is the part due to
anisotropic flow, and
$\delta_n$ represents the so called non-flow contribution,
that comes from correlations not related to the initial system geometry.
If non-flow is small, Eq.~(\ref{twoParticleFlowEstimate})
can be used to measure $v_n$, but in general the non-flow
contribution is not negligible.
To suppress non-flow one can exploit the collective nature 
of anisotropic flow using multi-particle correlations.   
The method based on multi-particle cumulants ({\it genuine}
multi-particle correlations) to measure anisotropic flow was proposed 
in~\cite{Borghini:2000sa,Borghini:2001vi,Borghini:2001zr,Adler:2002pu}.
This method allows to subtract non-flow effects 
from flow measurements order by order. 
Note that some experimental artifacts, such as
track splitting, in the analysis also contribute to the two particle
correlation; in this respect multi-particle techniques 
are also valuable, as they suppress such contributions as well.

One of the problems in using multi-particle correlations is the computing
power needed to go over all possible particle multiplets, which
practically prohibits calculations of correlations of order larger
than $k=3$ (three-particle correlations).
To avoid this problem, it was suggested in~\cite{Borghini:2000sa}
to express cumulants in terms of moments of the magnitude of the
corresponding flow vector $Q_n$, defined as: 
\begin{equation}
Q_{n} \equiv \sum_{i=1}^M 
e^{in\phi_i}\,,
\label{Qvector}
\end{equation}
where $M$ is the number of particles.
Unfortunately, flow estimates from cumulants constructed in such a way
were systematically biased by the interference between various harmonics.
An improved cumulant method using the formalism of generating
functions suggested in~\cite{Borghini:2001vi,Borghini:2001zr} 
fixed the problem of interfering  harmonics while keeping 
the number of operations still linear with multiplicity $M$.
For this approach the analytical calculations become
rather tedious and therefore the solutions are obtained using 
interpolation formulae. 
Unfortunately this introduces numerical uncertainties and requires
tuning of interpolating parameters for different  
values of the flow harmonics $v_n$ and multiplicity.
More recently a Lee-Yang-Zero's sum 
method~\cite{BBO:2003a,Bhalerao:2003xf,Borghini:2004ke,Bilandzic:2008nx}
has been developed to suppress non-flow contribution to all orders. 
Closely related to that are methods of Fourier and Bessel transforms
of the $Q$-distributions~\cite{Voloshin:2006gz}, and the method of direct
fitting of the $Q$-distribution. 
All these methods, while indeed being almost insensitive to non-flow, 
are biased by interference of different harmonics.

In this paper we present a new method to calculate multi-particle
cumulants in terms of moments of (in general, different harmonics)
$Q$-vectors.  
In our approach the cumulants are not biased by
interference between various harmonics, interpolating formulas
used in the formalism of generating functions are not
needed, and, moreover, all detector effects can be
disentangled from the flow estimates in
a single pass over the data at the level of or better than any other
  method. 
The number of operations required in our
approach  is still $\propto M$ for each $k$.
Since in our approach cumulants are calculated without any
approximation and  directly  from the data
we often call them {\it direct cumulants} 
(also referred to  as {\it $Q$-cumulants} because they are expressed
analytically in terms of different harmonic $Q$-vectors).  

Flow fluctuations are an important part of an anisotropic flow study. 
It is believed that flow fluctuations are mostly determined by
initial geometry fluctuations~\cite{Miller:2003kd}
 of the system created in a collision. 
An important consequence of this is that the anisotropic flow develops
relative to the so-called participant plane(s) instead of the reaction
plane determined by the direction of the impact 
parameter~\cite{Manly:2005zy}.
We note that the method  to calculate cumulants
proposed in this paper is not influenced by how exactly the anisotropic
flow is being developed.

In our simulations we show results obtained up to the 8-th order
cumulant, although we think that in practice there is little advantage
to go higher than order six, because going to higher order does not remove
the systematic uncertainty related to contribution from clusters 
exhibiting flow (see the discussion of systematic uncertainties
associated with cumulant analysis in~\cite{Adams:2003zg}).  
For example, in a 4-particle correlation analysis this bias corresponds
to the situation when two particles are correlated 
because they are coming from the same cluster and, in addition, 
correlated with another two particles via flow.

The paper is organized as follows. After the main definitions are
introduced in section~\ref{sec:defs}, 
we describe how the so-called {\em reference flow} can be
calculated. The reference flow is an average flow in
some momentum window; it is needed for the calculation of the differential
flow of particles of interest. To optimize the procedure, the
reference flow can be calculated using weights, e.g. weighted with
transverse momentum of the particle. Thus the reference flow can be
noticeably different from integrated flow of the same particles.
Section~\ref{sec:diffFlow} describes how the differential flow is calculated.
To show how the method works in different environments
and how it compares to some other methods we show simulation 
results in section~\ref{sec:simu}.
Finally, we summarize the main features of the method. 
Technical details, including the derivation of the main equations, 
equations in case of using non-unity weights in the calculation of 
reference flow, and acceptance effects are provided in Appendices.

\section{Multi-Particle azimuthal correlations and cumulants}
\label{sec:defs}

In this paper we discuss mostly 2- and 4-particle
azimuthal correlations (formulae for 6-particle correlation are
provided in the Appendix), 
but the generalization to azimuthal correlations
involving more particles is straightforward.
The method can be easily applied for calculations of {\it mixed
  harmonics} multi-particle correlations. In fact, mixed harmonics
correlations are needed in our approach for calculations of any
multi-particle correlations with order higher than 2.
Presenting 4-particle correlations below, we also show how the
3-particle correlations, involving one particle of a double harmonic 
can be calculated.
All the correlations are obtained by first averaging  over all
particles in a given event and then averaging over all events.
The latter may involve weights depending on event multiplicity. 

We define \textit{single-event} average 2- and
4-particle azimuthal correlations in the following way: 
\begin{eqnarray}
\mean{2} &\equiv&
\left< e^{in(\phi_{1} - \phi_{2})} \right> \equiv \frac{1}{P_{M,2}}\,
\sump_{i,j} e^{in(\phi_i-\phi_j)}\,,
\label{2pCorrelationSingleEvent}
\\ 
\mean{4} &\equiv& 
\la e^{in(\phi_{1} + \phi_{2}-\phi_{3}-\phi_{4})} \ra
\nonumber\\  &\equiv&
\frac{1}{P_{M,4}}\,\sump_{i,j,k,l} \,
e^{in(\phi_i+\phi_j-\phi_k-\phi_l)}\,,
\label{4pCorrelationSingleEvent} 
\end{eqnarray}
where $P_{n,m}=n!/(n-m)!$, and  
the prime in the sum $\sump$ means that all indices 
in the sum must be taken different.

The second step involves averaging over all events: 
\begin{eqnarray}
\dmean{2} &\equiv & \left<\!\left< e^{in(\phi_{1} -
  \phi_{2})} \right>\!\right>\nonumber\\ 
&\equiv&
\frac{\ds \sum_{\rm events} (W_{\left<2\right>})_i \left<2\right>_i}
{\ds \sum_{\rm events}
  (W_{\left<2\right>})_i}\,,
\label{2pCorrelationAllEvents}
\\
\dmean{4}&\equiv& 
\left<\!\left< e^{in(\phi_{1} + \phi_{2} - \phi_{3} 
- \phi_{4})} \right>\!\right>\nonumber\\
&\equiv&\frac{\ds \sum_{\rm events} 
(W_{\left<4\right>})_i \left<4\right>_i}
{\ds \sum_{\rm events} (W_{\left<4\right>})_i}\,,
\label{4pCorrelationAllEvents}
\end{eqnarray}
where by double brackets we denote an
average, first over all particles and then over all events. 
$W_{\left<2\right>}$ and $W_{\left<4\right>}$ are the event weights,
 which are used to minimize the effect of multiplicity variations
 in the event sample on the estimates of 2- and 4-particle correlations. 
In general, the optimal choice of weights 
would be determined by the multiplicity dependence of $v_n$.
The best approach might be 
to calculate the cumulants at fixed $M$
and then average over the entire event sample.   
In our calculations, with $v_n$ independent of multiplicity,
we use:
\begin{eqnarray}
W_{\left<2\right>}&\equiv& M(M-1)\,,\label{W<2>}\\
W_{\left<4\right>}&\equiv& M(M-1)(M-2)(M-3)\label{W<4>}\,.
\end{eqnarray}
The above choice for the event weights takes into account 
the number of different 2- and 4-particle combinations in an event
with multiplicity $M$.

The general formalism of cumulants 
was introduced into flow analysis by Ollitrault {\it et
  al}~\cite{Borghini:2000sa,Borghini:2001vi,Borghini:2001zr}. 
We will use below the notations from those papers.
The $2^{\rm nd}$ order cumulant, $c_{n}\{2\}$, is simply 
an average of 2-particle correlation defined in 
Eq.~(\ref{2pCorrelationAllEvents}): 
\begin{equation}
c_{n}\{2\} = \left<\left<2\right>\right>\,.
\label{cn2}
\end{equation}
As was pointed out first in~\cite{Borghini:2001vi} the
\textit{genuine} $4$-particle correlation (i.e. $4$-particle
cumulant), is given by:  
\begin{equation}
c_{n}\{4\} = \left<\left<4\right>\right> 
-2\cdot\left<\left<2\right>\right>^2\,.
\label{cn4}
\end{equation}
Expressions (\ref{cn2}) and (\ref{cn4}) are applicable only for
detectors with uniform acceptance and will be generalized in
Appendix~\ref{aNUA} to extend their applicability for 
detectors with non-uniform acceptance.

Different order cumulants provide
independent estimates for the same reference harmonic $v_n$. 
In particular~\cite{Borghini:2001vi}:  
\begin{eqnarray}
v_n\{2\} &=&\sqrt{c_n\{2\}}\,,\label{refFlowFromCumulants2nd}\\
v_n\{4\} &=&\sqrt[4]{-c_n\{4\}}\,,\label{refFlowFromCumulants4th}
\end{eqnarray}
where the notation $v_n\{2\}$ is used to denote the reference flow
$v_n$ estimated from the $2^{\rm nd}$ order cumulant
$c_{n}\{2\}$, and $v_n\{4\}$ stands for the reference flow 
$v_n$ estimated from the $4^{\rm th}$ order cumulant $c_{n}\{4\}$.

\section{Reference flow}
\label{sec:refFlow}

To obtain the $2^{\mathrm{nd}}$ order cumulant it suffices to
separate diagonal and off-diagonal terms in $\left|Q_n\right|^2$: 
\begin{equation}
\left|Q_n\right|^2 = \sum_{i,j=1}^{M} e^{in(\phi_i-\phi_j)}\,
= M+\sump_{i,j}e^{in(\phi_i-\phi_j)}\,,
\label{|Q_n|^2}
\end{equation}
which can be trivially solved to obtain $\left<2\right>$:
\begin{equation}
\left<2\right> = \frac{\left|Q_n\right|^2-M}{M(M-1)}\,.
\label{2p:result}
\end{equation}
The event averaging is being performed via
Eq. (\ref{2pCorrelationAllEvents}). The resulting expression for
$\dmean{2}$ is than used to estimate $2^{\rm nd}$ order cumulant (see
Eq. (\ref{cn2})), which in turn is used to estimate the reference flow
harmonic $v_n$ by making use of Eq.~(\ref{refFlowFromCumulants2nd}). 
 
To obtain the $4^{\mathrm{th}}$ order cumulant we start with 
the decomposition of  $\left|Q_n\right|^4$ (for details, see
Appendix~\ref{sec:appEqs}) 
\begin{equation}
\left|Q_n\right|^4 = Q_n Q_n Q_n^* Q_n^* =
\sum_{i,j,k,l=1}^{M}e^{in(\phi_i+\phi_j-\phi_k-\phi_l)}\,.
\end{equation}
We have four distinct cases for the indices
$i$, $j$, $k$ and $l$: 1) they are all different (4-particle
correlation), 
2) three are different, 3) two are different or 4) they are all the same. 
Note, that the case of three different indices 
corresponds to the so-called mixed harmonics 3-particle correlations,
in many analyses of great interest by 
itself~\cite{Adams:2003zg,Borghini:2002vp}.
Equations for 3-particle correlations are provided in
Appendix~\ref{sec:appEqs}. 
Taking everything into account, we obtain
the following analytic result for the
single-event average 4-particle correlation defined in
Eq. (\ref{4pCorrelationSingleEvent}): 
\begin{eqnarray}
\left<4\right>&=&\frac{\left|Q_n\right|^4+
\left|Q_{2n}\right|^2 -2\cdot
\mathfrak{Re}\left[Q_{2n}Q_n^*Q_n^*\right]}{M(M-1)(M-2)(M-3)}
\nonumber\\
&-&2\,\frac{2(M-2)\cdot\left|Q_n\right|^2-M(M-3)}{M(M-1)(M-2)(M-3)}\,.
\label{4p:result}
\end{eqnarray}
The reason why the originally proposed cumulant analysis~\cite{Borghini:2000sa}
was biased lies in the fact that 
the terms consisting of $Q$-vectors evaluated in {\it
  different} harmonics (for instance terms $\left|Q_{2n}\right|^2$ and
$\mathfrak{Re}\left[Q_{2n}Q_{n}^*Q_{n}^*\right]$) 
have been neglected. As seen from
Eq. (\ref{4p:result}), such terms do appear in the analytic results
and are crucial in disentangling the interference between harmonics.  In particular, if a higher harmonic $v_{2n}$ is present than $\left|Q_{n}\right|^4$ picks up an additional contribution 
depending on that harmonic, namely $v_{2n}^2 M(M\!-\!1)\! +\! v_n^2v_{2n} 2M(M\!-\!1)(M\!-\!2)$, which is exactly 
canceled out with the contribution of harmonic $v_{2n}$ to $\left|Q_{2n}\right|^2$ and $\mathfrak{Re}\left[Q_{2n}Q_n^*Q_n^*\right]$, 
which read $ Mv_{2n}^2(M\!-\!1)$ and $M(M\!-\!1)(M\!-\!2)v_{n}^2v_{2n}\!+\!M(M\!-\!1)v_{2n}^2$, respectively. 

The final, event averaged 4-particle azimuthal
correlation, $\left<\left<4\right>\right>$, is then obtained by making
use of Eqs. (\ref{4pCorrelationAllEvents}) and (\ref{W<4>}). 
Using  $\left<\left<4\right>\right>$ and
$\left<\left<2\right>\right>$ one can calculate the
$4^{\mathrm{th}}$ order cumulant from Eq. (\ref{cn4}).

The reference flow is mainly used to calculate differential flow. 
Therefore, one can optimize the calculation
of reference flow to minimize the uncertainties in the final results. 
This is done by using different weights
(e.g. particle transverse momentum) in the
definition of flow vectors used in reference flow calculations.
We provide all the equations necessary for calculations with weights
in Appendix~\ref{aPW}.

The equations so far are applicable for an analysis with a 
detector with full uniform azimuthal coverage. 
In a non-ideal case one needs to take into account the acceptance
corrections~\cite{Bhalerao:2003xf,Selyuzhenkov:2007zi}.
Acceptance affects the cumulants in three ways: 
(i) contributions from additional terms, e.g. proportional to
$\left<\left< \cos n\phi \right> \right>$
or $\left<\left< \sin n\phi \right> \right>$,
that for a detector with full uniform azimuthal coverage are identical
 to zero, 
(ii) contributions from other flow harmonics, and 
(iii) the cumulant might be rescaled, which at the end can
affect the final extracted flow values. 
We refer to Refs.~\cite{Bhalerao:2003xf,Selyuzhenkov:2007zi} 
for a more complete discussion of acceptance effects.
In practice the most important correction is the first one, 
for which we provide the full
set of equations for a 2- and 4- particle cumulant analysis. 

The generalized $2^{\rm nd}$ order cumulant 
which can also be used for detectors with non-uniform acceptance is: 
\begin{eqnarray}
c_{n}\{2\} &=&\left<\left<2\right>\right>
-\mathfrak{Re}\bigg\{
\big[\left<\left<\cos n\phi_1\right>\right>
+i\left<\left<\sin n\phi_1\right>\right>\big]\nonumber\\
&&\times\big[\left<\left<\cos n\phi_2\right>\right>
-i\left<\left<\sin n\phi_2\right>\right>\big]\bigg\}\nonumber\\
&=&\left<\left<2\right>\right> - \left<\left<\cos
  n\phi_1\right>\right>^2
-\left<\left<\sin n\phi_1\right>\right>^2\,,
\label{nuaHere}
\end{eqnarray}
where for the last line we have used the fact that for
instance $\left<\left<\cos n\phi_1\right>\right>$ and  
$\left<\left<\cos n\phi_2\right>\right>$ are the same quantities apart
from the trivial relabeling. Remarkably, only two additional terms
appear in Eq.~(\ref{nuaHere}), namely $\left<\left<\cos
n\phi_1\right>\right>^2$ and $\left<\left<\sin
n\phi_1\right>\right>^2$, which counterbalance the bias to
$\left<\left<2\right>\right>$ coming from very general 
detector inefficiencies.
Further details on treating the acceptance effects, including formulae for
the $4^{\rm th}$ order cumulant are provided in Appendix~\ref{aNUA}.

\section{Differential flow}
\label{sec:diffFlow}

Once the reference flow has been estimated with the help of the formalism
from previous section, we proceed to  the calculation of differential flow.
For that, all particles selected for flow analysis are labeled as 
{\it Reference Flow Particle},  RFP, and/or {\it Particle Of Interest}, POI.
These labels are needed because flow analysis is being
performed in two steps. In the first step one estimates the
reference flow by using only the RFPs, while in the second step we
estimate the differential flow of POIs with respect to the reference
flow of the RFPs obtained in the first step. 

\subsection{Reduced multi-particle azimuthal correlations }
\label{sRMAC}

For {\it reduced} single-event average 2- and 4-particle azimuthal
correlations we use the following notations and definitions: 
\begin{eqnarray}
\left<2'\right>&\equiv&\left< e^{in(\psi_{1} - \phi_{2})} \right>
\nonumber\\
&\equiv&\frac{1}{m_pM\!-\!m_q}\,
\sum_{i=1}^{m_{p}}\sump_{j=1}^{M}e^{in(\psi_i-\phi_j)}\,, 
\label{2pReducedCorrelationSingleEvent}\\ 
\left<4'\right>&\equiv&\left< e^{in(\psi_{1} +
  \phi_{2}-\phi_{3}-\phi_{4})} \right>
\nonumber\\ 
&\equiv&\frac{1}{(m_pM\!-\!3 m_q)(M\!-\!1)(M\!-\!2) } 
\nonumber\\
&\times&\sum_{i=1}^{m_p}\,\sump_{j,k,l=1}^{M} 
e^{in(\psi_i+\phi_j-\phi_k-\phi_l)}\,,
\label{4pReducedCorrelationSingleEvent}
\end{eqnarray}
where $m_p$ is the total number of particles labeled as POI (some of
which might have been also labeled additionally as RFP), 
$m_q$ is the total number of particles labeled
{\it both} as RFP and POI, 
$M$ is the total number of particles labeled as RFP (some of which
might have been 
also labeled additionally as POI) in the  event,
$\psi_i$ is the azimuthal angle of the $i$-th particle labeled as POI
and taken from the phase window of interest (taken even if it was also
additionally labeled as RFP), $\phi_j$ is the azimuthal angle of the
$j$-th particle labeled as RFP (taken
even if it was also additionally labeled as POI).  
$\sump$, as before, denotes the sum with all indices taken different.

Final, event averaged reduced 2- and
4-particle correlations are given by: 
\begin{eqnarray}
\dmean{2'} &\equiv & \frac{\ds \sum_{\rm events} (w_{\left<2'\right>})_i 
\mean{2'}_i}{\ds \sum_{\rm events} (w_{\left<2'\right>})_i}\,,
\label{2pReducedCorrelationAllEvents}\\
\dmean{4'}&\equiv&\frac{\ds \sum_{\rm events} (w_{\left<4'\right>})_i 
\left<4'\right>_i}{\ds \sum_{events} (w_{\left<4'\right>})_i}\,.
\label{4pReducedCorrelationAllEvents}
\end{eqnarray}
In our calculations we use
 event weights $w_{\left<2'\right>}$ and
$w_{\left<4'\right>}$ defined as: 
\begin{eqnarray}
w_{\left<2'\right>}&\equiv&m_pM-m_q\,,\label{W<2'>}\\
w_{\left<4'\right>}&\equiv&(m_pM-3 m_q)(M-1)(M-2)\label{W<4'>}\,.
\end{eqnarray}
%

\subsection{Differential cumulants}
\label{ssDQC}

We derive equations for the differential equations with the help of
$p$- and $q$-vectors; the former built out of all POIs ($m_p$ in
total), and the second only from POI labeled also as RFP ($m_q$ in total):
\begin{eqnarray}
p_{n} &\equiv& \sum_{i=1}^{m_p} e^{in\psi_i}\,,
\label{p-vectorDefinitionUnitWeight}
\end{eqnarray}
\begin{eqnarray}
q_{n} &\equiv& \sum_{i=1}^{m_q} e^{in\psi_i}\,.
\label{q-vectorDefinitionUnitWeight}
\end{eqnarray}
The $q$-vector is introduced here in order to subtract effects of autocorrelations. 
Using those, we have obtained the following equations for the
average reduced single- and all-event 2-particle correlations: 
\be
\left<2'\right> &=& \frac{p_{n} Q_n^* - m_q}{m_p M\!-\!m_q}\,,
\label{2PrimeSingleEvent:Result}\\
\left<\left<2'\right>\right> &=& 
\frac{\sum_{i=1}^N (w_{\left<2'\right>})_i \left<2'\right>_i}
{\sum_{i=1}^N (w_{\left<2'\right>})_i}\,.
\label{2PrimeAllEvent:Result}
\ee

For detectors with uniform azimuthal acceptance 
the differential $2^{\rm nd}$ order cumulant is given by 
\begin{equation}
d_n\{2\} = \left<\left<2'\right>\right>\,,
\label{qc2nd:Definition:Diff}
\end{equation}
where, again we use notation from Ref.~\cite{Borghini:2001vi}.
We present equations for the
case of detectors with non-uniform acceptance in Appendix~\ref{aNUA}.

Estimates of differential flow 
$v'_n$ are being denoted as $v'_n\{2\}$ and are given by
\cite{Borghini:2001vi}: 
\begin{eqnarray}
v'_n\{2\}&=& \frac{d_n\{2\}}{\sqrt{c_n\{2\}}}\,. 
\label{diffFlow2nd:Result}
\end{eqnarray}

Below we present the corresponding formulae for reduced
 4-particle correlations:
\begin{eqnarray}
\left<4'\right> &=& \bigg[
  p_{n}Q_{n}Q_{n}^{*}Q_{n}^{*}-q_{2n}Q_{n}^{*}Q_{n}^{*}-p_{n}Q_{n}Q_{2n}^{*}
  \nonumber\\
&-&2\cdot M p_{n}Q_{n}^{*}-2\cdot m_q \left|Q_{n}\right|^2+7\cdot
  q_{n}Q_{n}^{*} 
\nonumber\\
&-& Q_{n}q_{n}^{*}+q_{2n}Q_{2n}^{*}+2\cdot p_{n}Q_{n}^{*}
\nonumber\\
&+&2\cdot m_q M - 6\cdot m_q\bigg]
\nonumber\\
&/&\bigg[(m_p M -3m_q)(M-1)(M-2)\bigg]\,,
\label{4PrimeSingleEvent:Result}\\
\left<\left<4'\right>\right>&=&\frac{\sum_{i=1}^N
  (w_{\left<4'\right>})_i \left<4'\right>_i}{\sum_{i=1}^N
  (w_{\left<4'\right>})_i}\,.
\label{4PrimeAllEvent:Result}
\label{eventWeightsFor4Prime}
\end{eqnarray}
The $4^{\rm th}$ order differential cumulant is given 
by~\cite{Borghini:2001vi}:  
\begin{eqnarray}
d_n\{4\}&=& \left<\left<4'\right>\right>-2\cdot\left<\left<2'\right>\right>
\left<\left<2\right>\right>
\label{diffCumulant4th}\,.
\end{eqnarray}
Equations for the
case of detectors with non-uniform acceptance are again presented in
Appendix~\ref{aNUA}. 

Having obtained estimates for $d_n\{4\}$ and
$c_n\{4\}$, we can estimate differential flow~\cite{Borghini:2001vi}:   
\begin{eqnarray}
v'_n\{4\}&=& -\frac{d_n\{4\}}{(-c_n\{4\})^{3/4}} 
\label{diffFlow4th:Result}\,. 
\end{eqnarray}
Similarly to reference flow, we use the notation $v'_n\{4\}$ for differential
flow harmonics $v'_{n}$ obtained from $4^{\rm th}$ order cumulants. 
$v'_n\{4\}$ and $v'_n\{2\}$ are independent
estimates for the same differential flow harmonic $v'_{n}$.

\section{Simulation results}
\label{sec:simu}

We have tested the new method with extensive simulations. 
The results, presented below, show that the method effectively
suppresses non-flow contributions, 
illustrate the ability to remove the interference of the different 
harmonics, show the applicability for detectors having significant
acceptance ``holes'', and give an example of a differential flow analysis.
In the figures, $v_2\{\rm MC\}$, shown  
in the first bin, represents the Monte Carlo estimate for
$v_n$, which was obtained using the known
reaction plane event-by-event. 
Other estimates in the figures are obtained without using this information. 

\begin{figure}[thb]
 \begin{center}
   \includegraphics[width=0.5\textwidth]{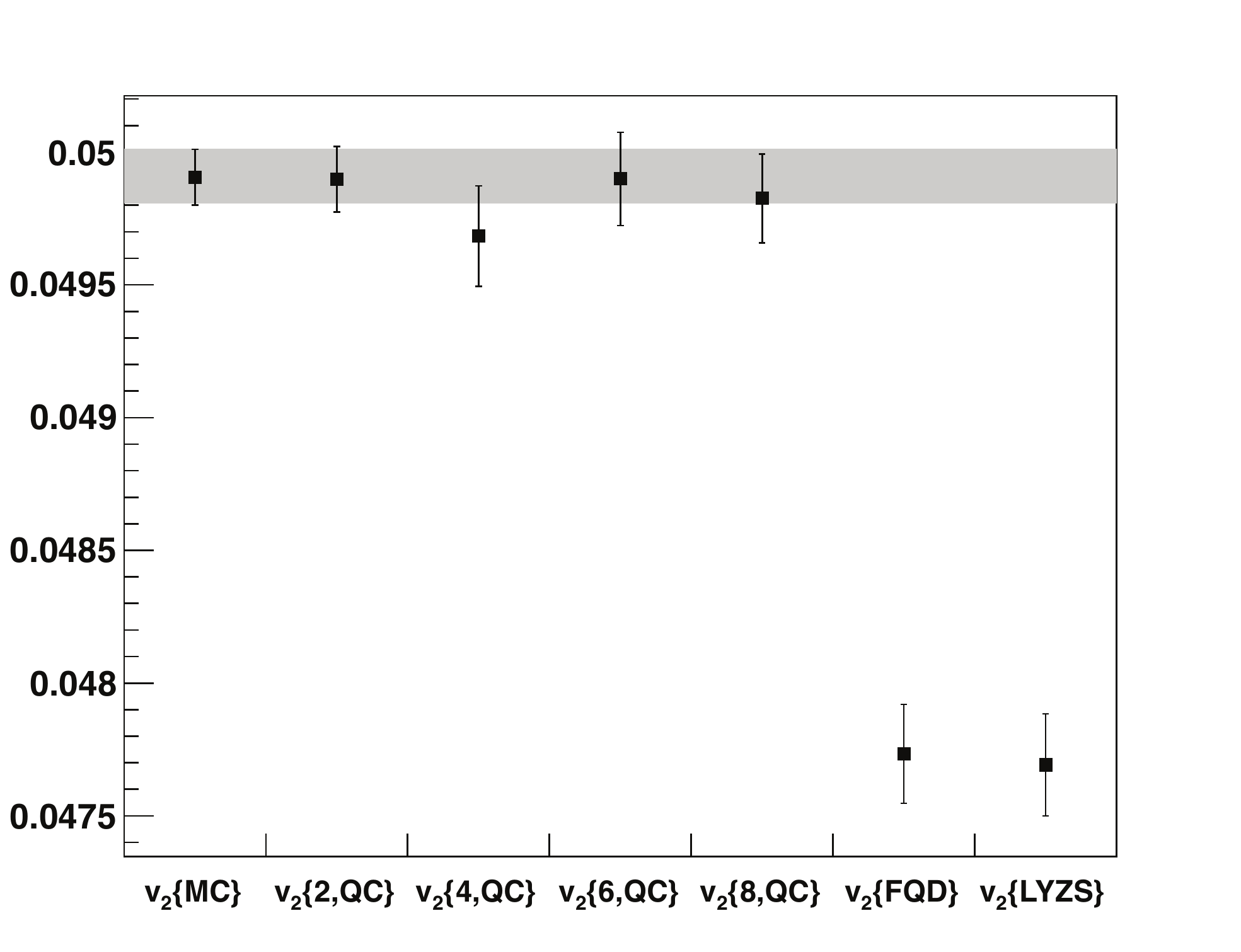}
   \caption{
     Elliptic flow extracted by different methods for
     $10^5$ simulated events with multiplicity $M=500$, 
     $v_2=0.05$ and at the same time $v_4=0.1$.
   }
    \label{fig:interference}
 \end{center}
\end{figure}
Figure~\ref{fig:interference} shows the results from a simulation
of events with anisotropic flow present in two harmonics, the second and
the fourth. 
Elliptic flow estimated by different methods is shown in the figure. 
A clear bias is observed in the estimates from fitting of the
$Q$-distribution method and the Lee-Yang Zero's Sum method, 
labeled as $v_2\{\rm FQD\}$ and $v_2\{\rm LYZS\}$, respectively. 
Results obtained with direct cumulants of different order, 
labeled as $v_2\{k,{\rm QC}\}$, are unaffected by
  $v_4$ interference. 

\begin{figure}[thb]
 \begin{center}
   \includegraphics[width=0.45\textwidth]{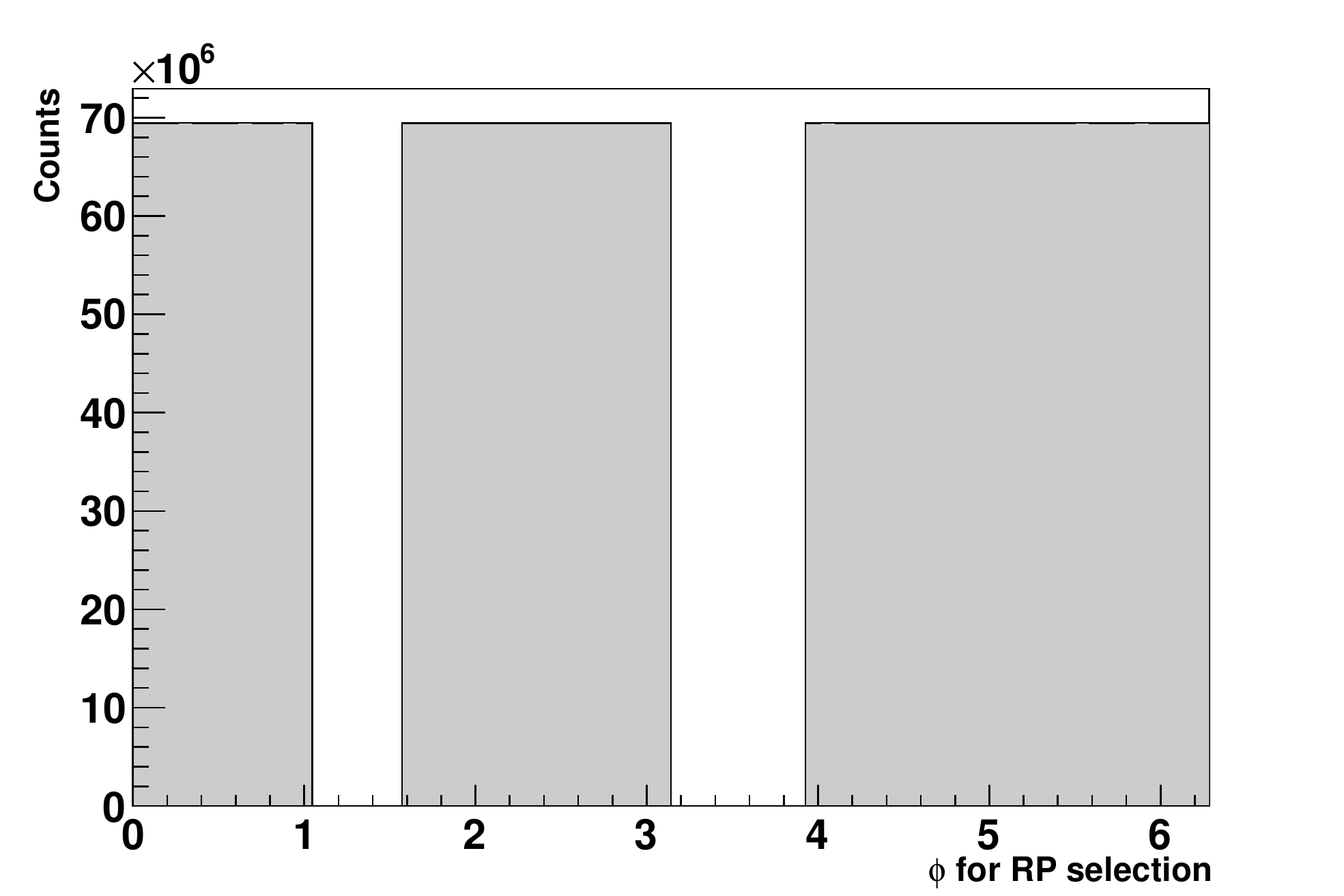}
   \centerline{ (a) }
   \includegraphics[width=0.45\textwidth]{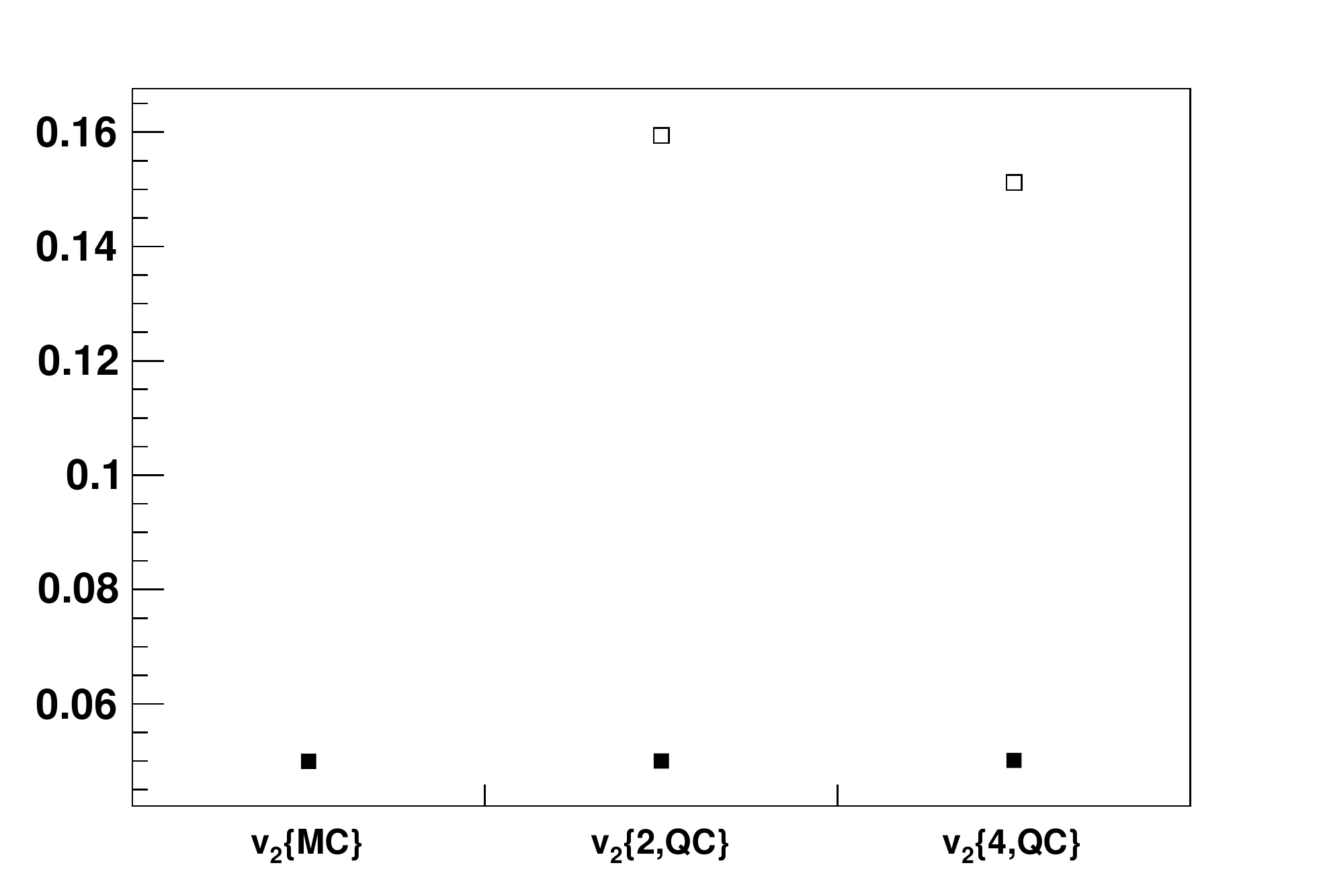}
   \centerline{ (b) }
   \includegraphics[width=0.45\textwidth]{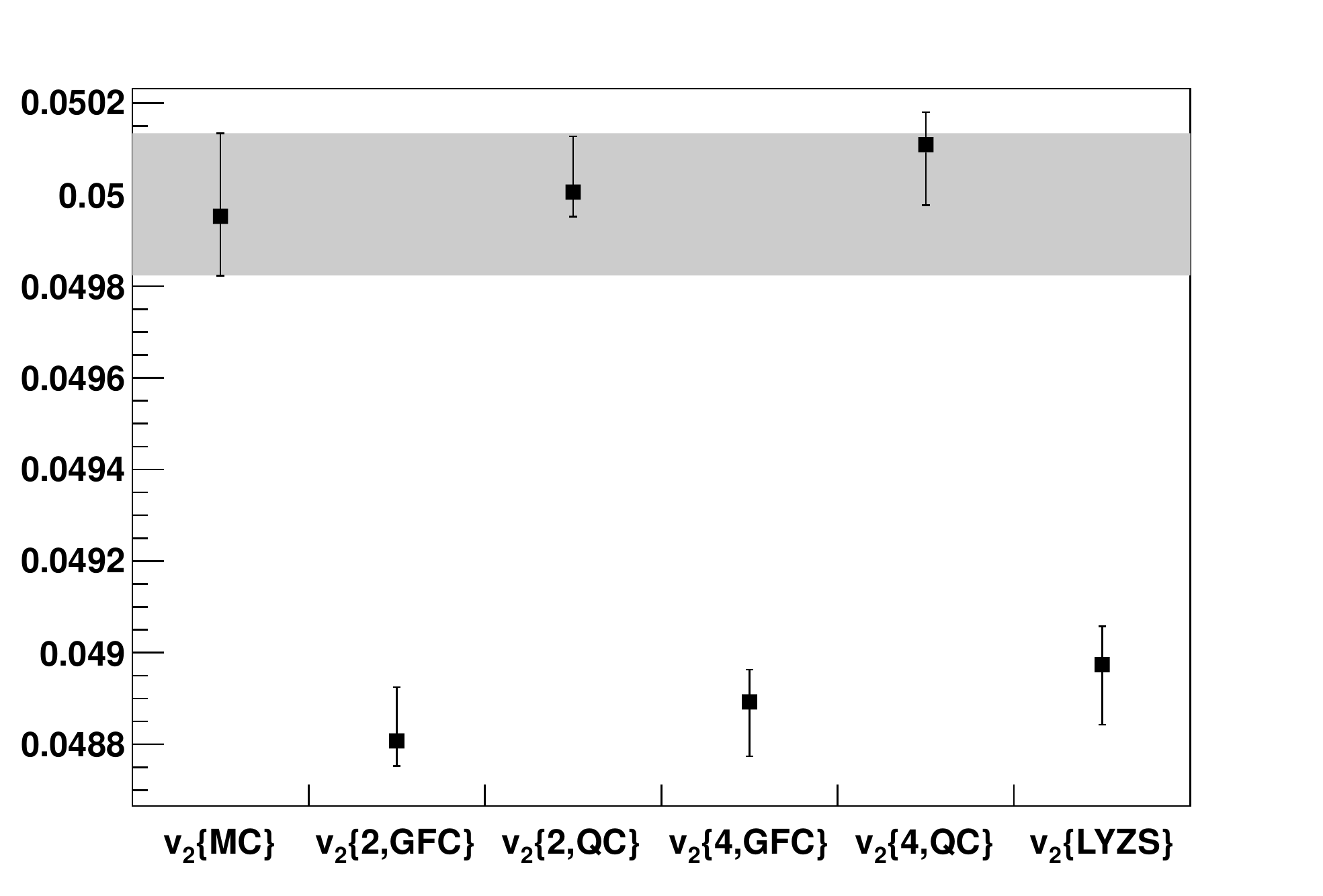}
   \centerline{ (c) }
   \caption{
     a) The azimuthal distribution of accepted particles.
     b) Extracted elliptic flow accounting for acceptance
    effects, closed markers, and without, open markers.
    c) Extracted elliptic flow accounting for acceptance
    effects in different methods.
}
    \label{fNUA1}
 \end{center}
\end{figure}
To demonstrate that the method works well even in cases with rather bad
acceptance we simulated $10^7$ events with $v_2=0.05$ 
for a detector that had two large ``holes'' 
(see  Fig.~\ref{fNUA1}a). 
Figure~\ref{fNUA1}b shows
the obtained $v_2$ estimates using Eqs. (\ref{cn2}) and (\ref{cn4}) 
which are valid for detectors with perfect acceptance using open markers. 
Clearly these values are strongly biased.
The $v_2$ estimates obtained from the more general equations 
for cumulants, namely Eqs. (\ref{gen2ndQC}) and (\ref{gen4thQC}), 
which do account for the acceptance effects are shown as closed
markers and agree with the Monte Carlo estimate.
In Fig.~\ref{fNUA1}c we look in more detail at the agreement with 
the Monte Carlo estimate and, in addition, compare to other methods.

The figure clearly shows that detector effects are corrected for at
 the level of or better than other methods.

\begin{figure}[thb]
 \begin{center}
   \includegraphics[width=0.5\textwidth]   {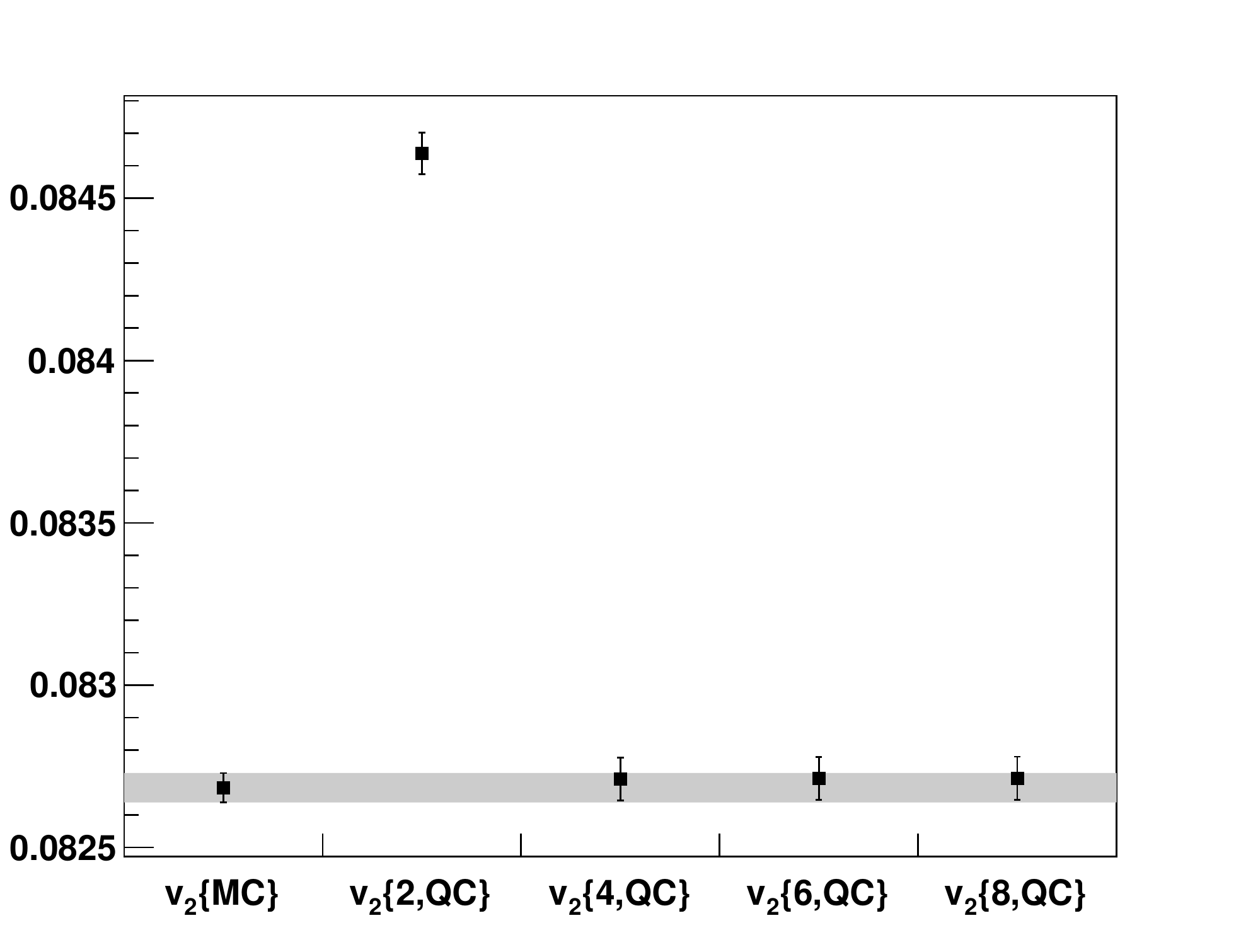}
   \caption{
    Reference flow extracted from particles labeled as RFPs 
    (pions in Therminator) 
   }
    \label{referenceFlowPions}
 \end{center}
\end{figure}

\begin{figure}[thb]
 \begin{center}
   \includegraphics[width=0.5\textwidth]   {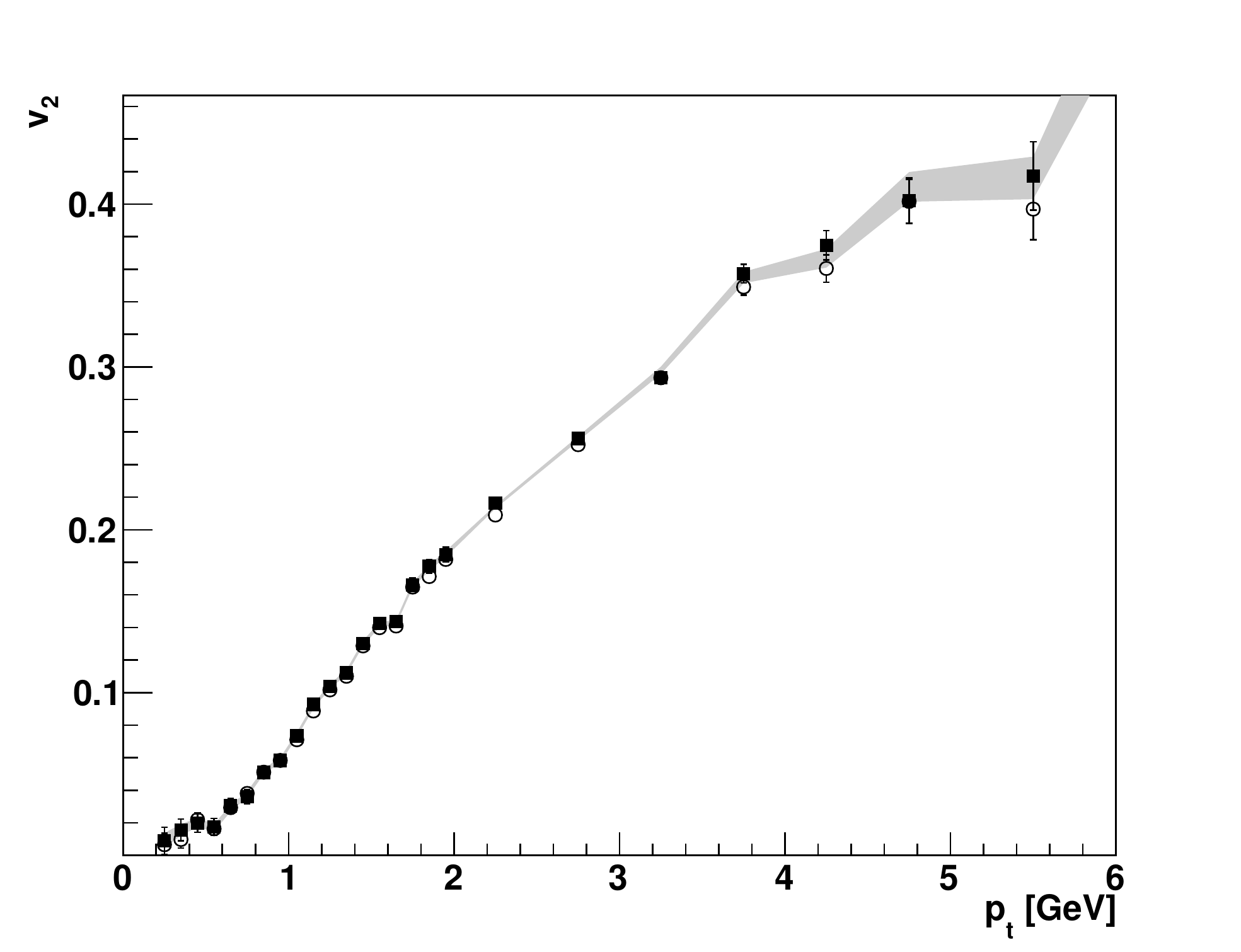}
   \caption{
   Differential flow extracted for particles labeled as POIs 
   from Therminator events (in this example we used protons). 
   The open circles denote $2^{\rm nd}$ order estimate
   (Eq. (\ref{diffFlow2nd:Result})) and closed squares denote $4^{\rm
   th}$ order estimate (Eq. (\ref{diffFlow4th:Result})).} 
    \label{diffFlowProtons}
 \end{center}
\end{figure}

\begin{figure}[thb]
 \begin{center}
   \includegraphics[width=0.5\textwidth] {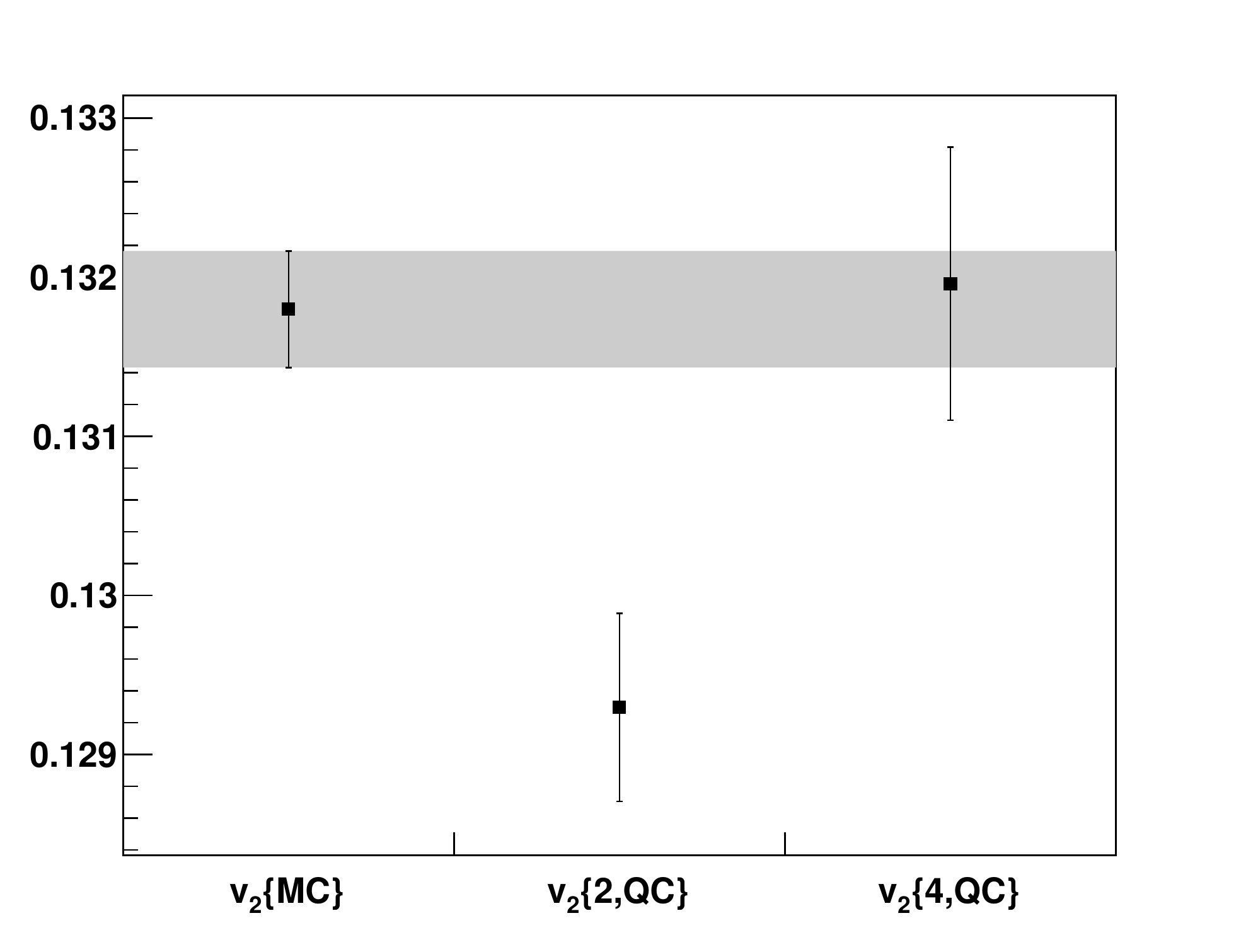}
   \caption{$p_t$-integrated flow calculated from
   Eq. (\ref{integratedFlow}) of protons whose differential flow is
   presented in Fig.~\ref{diffFlowProtons}.
   } 
    \label{integratedFlowProtons}
 \end{center}
\end{figure}

As an example of a differential flow analysis we show
results for $v'_2(p_t)$ obtained with Therminator~\cite{Kisiel:2005hn}. 
As RFPs we select pions and as POIs we select protons. 
In the first step we estimate the reference flow by only making 
use of particles labeled as RFPs (using Eqs.~(\ref{cn2}), (\ref{cn4}),
(\ref{refFlowFromCumulants2nd}) and (\ref{refFlowFromCumulants4th})). 
The estimates of reference flow are presented 
in Fig.~\ref{referenceFlowPions}. 
In the second step we estimate the differential flow of POIs (in this
example protons were labeled as POIs) with respect to the reference
flow of RFPs estimated in the first step. 
For each $p_t$ bin we evaluate $d_{n}\{2\}$ and $d_{n}\{4\}$, 
and use equations (\ref{diffFlow2nd:Result}) and (\ref{diffFlow4th:Result}) 
to estimate differential flow. 
The differential flow results for protons are presented
in Fig.~\ref{diffFlowProtons}. 
The resulting $p_t$-integrated flow of protons calculated by making 
use of Eq.~(\ref{integratedFlow}) is presented in
Fig.~\ref{integratedFlowProtons}.  
The figures for the integrated flow of the RFPs and POIs clearly 
show that the $2^{\rm nd}$ order cumulant is biased by nonflow 
while the higher order cumulants are in perfect agreement with
the Monte Carlo.

\section{Summary}
\label{sec:summary}

In summary, 
we propose a new method to calculate multi-particle azimuthal
correlations, which provides fast (in a single scan over the data) and
exact (no approximations) non-biased (no interference between
different harmonics) estimates for cumulants.
In the paper, we provide the corresponding formulae for correlations up to
the 6-th order, but the method, if needed, can be generalized for
higher orders.  

We have not discussed issues of the cumulant approach in general, such as
multiplicity fluctuations, flow fluctuations, and low
sensitivity for small flow values, but believe that our method will be
helpful in investigating all these questions.

The proposed method has been extensively tested in simulations and 
has been used for real data analysis by
the STAR and ALICE Collaborations~\cite{Dhevan:PhD,Ante:PhD}. 
Further details about the method, including equations 
for 8-particle correlations, equations for estimates and evaluation 
of statistical errors, comparison to other methods, 
can be found in~\cite{Ante:PhD}.

\acknowledgments{
We thank Dhevan Gangadharan, Rene Kamermans, 
Naomi~van~der~Kolk, Paul Kuijer, Mikolaj Krzewicki, Art Poskanzer, 
Gerard Smit, Paul Sorensen, Aihong Tang, Jim Thomas, Adam Trzupek, Fuqiang Wang and Evan Warren  
for their help, discussions, and interest in this work.
The work of SV was supported in part by the US Department of Energy, 
Grant No. DE-FG02-92ER40713. The work of AB and RS was supported in part by 
the Dutch funding agencies FOM and NWO.
}

\appendix
\section{Equations for 3-, 4- and 6- particle correlations 
\label{sec:appEqs}}

Below we use the following definitions: 
\begin{eqnarray}
\mean{2}\equiv\mean{2}_{n|n}&\!\equiv&\!
\frac{1}{P_{M,2}}\,\sump_{i,j=1}^{M}e^{in(\phi_i-\phi_j)}\,,
\label{2-p}\\
\mean{2}_{2n|2n}&\!\equiv&\!\frac{1}{P_{M,2} }\,\
\sump_{i,j=1}^{M}e^{i2n(\phi_i-\phi_j)}\,,
\label{2-p2n}\\
\left<3\right>_{2n|n,n}&\!\equiv&\!\frac{1}{P_{M,3} }\,\
\sump_{i,j,k=1}^{M} \!\!e^{in(2\phi_i-\phi_j-\phi_k)}\,,\\
\left<3\right>_{n,n|2n}&\!\equiv&\!\left<3\right>_{2n|n,n}^*\,,\\
\left<4\right> \equiv \left<4\right>_{n,n|n,n}&\!\!\!\equiv&\!\!
\frac{1}{P_{M,4}} \sump_{i,j,k,l=1}^{M}\!\!\!\! 
e^{in(\phi_i+\phi_j-\phi_k-\phi_l)}.
\label{4-p}
\end{eqnarray}
Using this notation one finds:
\begin{eqnarray}
\left|Q_n\right|^4 &=& \left<4\right>_{n,n|n,n}\cdot P_{M,4}
\nonumber\\
&+&\left[\left<3\right>_{2n|n,n}+\left<3\right>_{n,n|2n}\right] \cdot P_{M,3}
\nonumber\\
&+&\mean{2}_{n|n}\cdot 4 P_{M,2} (M-1) 
\nonumber\\
&+&\mean{2}_{2n|2n}\cdot P_{M,2}\nonumber\\
&+& 2 P_{M,2} + M\,.
\label{4h}
\end{eqnarray}
The 2-particle correlations $\mean{2}_{n|n}$ was already expressed in
terms of the $Q$-vector evaluated in harmonic $n$, see
Eq.~(\ref{2p:result}):
\begin{equation}
\mean{2}_{2n|2n} = \frac{\left|Q_{2n}\right|^2-M}{P_{M,2}}\,.
\label{2p2n2n}
\end{equation}
To obtain
$\left<3\right>_{2n|n,n}$ and $\left<3\right>_{n,n|2n}$ we have to 
decompose 
\begin{eqnarray}
Q_{2n}Q_n^*Q_n^* &=& \left<3\right>_{2n|n,n}\!\cdot\!
P_{M,3}\!+\!\mean{2}_{n|n}\!\cdot\! 2P_{M,2}
\nonumber\\
&+&\mean{2}_{2n|2n}\!\cdot\! P_{M,2}\!+\!1\cdot M\,,
\end{eqnarray}
and $Q_nQ_nQ_{2n}^*$.
After inserting results for $\mean{2}_{n|n}$ and
$\mean{2}_{2n|2n}$ given in 
Eqs. (\ref{2p:result}) and (\ref{2p2n2n}), 
we arrive at the following equality:
\begin{eqnarray}
\left<3\right>_{n,n|2n}+\left<3\right>_{2n|n,n}&=&2\,
\frac{\mathfrak{Re}\left[Q_{2n}Q_n^*Q_n^*\right]-
  2\cdot\left|Q_n\right|^2}{M(M-1)(M-2)}
\nonumber\\
&-&2\,\frac{\left|Q_{2n}\right|^2-2M}{M(M-1)(M-2)}\,.
\label{3p}
\end{eqnarray}
After inserting Eqs. (\ref{2p:result}), (\ref{2p2n2n}) and
(\ref{3p}) into Eq.~(\ref{4h}) and solving the resulting expression
for $\left<4\right>_{n,n|n,n}$ the single-event average 4-particle 
correlations (Eq.(\ref{4p:result})) follows. 

This derivation can be generalized to obtain 
analytic results for any higher order multi-particle azimuthal
correlations.  
Below we provide the expression for the 6-particle correlation: 
\begin{eqnarray}
\left<6\right>&\equiv&\frac{1}{P_{M,6}}
\sump_{i,j,k,l,m,n=1}^{M}
\!\!\!\!\!\! e^{in(\phi_i+\phi_j+\phi_k-\phi_l-\phi_m-\phi_n)}
\nonumber\\
&=&\frac{\left|Q_n\right|^6\!
+\!9\cdot\left|Q_{2n}\right|^2\left|Q_{n}\right|^2\!
-\!6\cdot\mathfrak{Re}\left[Q_{2n}Q_{n}Q_{n}^*Q_{n}^*Q_{n}^*\right]}
{M(M-1)(M-2)(M-3)(M-4)(M-5)}
\nonumber\\
&+&4\,\frac{\mathfrak{Re}\left[Q_{3n}Q_{n}^*Q_{n}^*Q_{n}^*\right]
-3\cdot\mathfrak{Re}\left[Q_{3n}Q_{2n}^*Q_{n}^*\right]}
{M(M-1)(M-2)(M-3)(M-4)(M-5)}
\nonumber\\
&+&2\,\frac{9(M-4)\cdot\mathfrak{Re}\left[Q_{2n}Q_{n}^*Q_{n}^*\right]
+2\cdot\left|Q_{3n}\right|^2}{M(M-1)(M-2)(M-3)(M-4)(M-5)}
\nonumber\\
&-&9\,\frac{\left|Q_{n}\right|^4+\left|Q_{2n}\right|^2}
{M(M-1)(M-2)(M-3)(M-5)}
\nonumber\\
&+&18\,\frac{\left|Q_{n}\right|^2}{M(M-1)(M-3)(M-4)}
\nonumber\\
&-&\frac{6}{(M-1)(M-2)(M-3)}\,.
\label{6th:results}
\end{eqnarray}
With that, the $6^{\rm th}$ order cumulant is given by
\begin{equation}
c_n\{6\}= \left<\left<6\right>\right>
-9\cdot\left<\mean{2}\right>\left<\left<4\right>\right>
+12\cdot\left<\mean{2}\right>^3\,,\label{6th}
\end{equation}
and the reference flow $v_n$ is estimated as 
\begin{equation}
v_{n}\{6\} = \sqrt[6]{\frac{1}{4}c_n\{6\}}\,.
\end{equation}
%

\section{Particle weights \label{aPW}}

Below we provide formulae to use for the case when the reference flow is
calculated using particle weights. 
For that we introduce a weighted $Q$-vector evaluated
in harmonic $n$: 
\begin{equation}
Q_{n,k} \equiv \sum_{i=1}^M w_i^k e^{in\phi_i}\,,
\label{weightedQvector}
\end{equation}
where $w_i$ is a particle weight of the $i$-th particle labeled as RFP
and $M$ is the total number of RFPs in an event. 
In general, we will need flow vectors with power $k$ up to 
the order of multi-particle correlations. 
Similarly, we define
\begin{eqnarray}
p_{n,k} &\equiv& \sum_{i=1}^{m_p} w_i^k e^{in\psi_i}\,.
\label{p-vectorDefinition}
\end{eqnarray}
Note that only particles which have a RFP label, have a non-unit
weight, while for the particles labeled
as POI \textit{only}, $w_i=1$.
For the subset of POIs which
consists of all particles labeled \textit{both} as POI and RFP ($m_q$ in
total) we introduce 
\begin{eqnarray}
q_{n,k} &\equiv& \sum_{i=1}^{m_q} w_i^k e^{in\psi_i}\,.
\label{q-vectorDefinition}
\end{eqnarray}

For RFPs we also introduce: 
\begin{eqnarray}
S_{p,k} &\equiv& \left[\sum_{i=1}^{M} w_i^k \right]^p\,,
\label{Spk}\\
\mathcal{M}_{abcd\cdots} &\equiv&
\!\!\!\!\!\sump_{i,j,k,l,\ldots=1}^{M}\!\!\!\!\!w_i^a w_j^b w_k^c
w_l^d \cdots \,. 
\label{Mabcd}
\end{eqnarray}
For all particles labeled \textit{both} as RFP and POI we evaluate
the following quantity: 
\begin{eqnarray}
s_{p,k} &\equiv& \left[\sum_{i=1}^{m_q} w_i^k \right]^p
\label{spk}\,,
\end{eqnarray}
while in the definition below the first sum runs over all POIs
in the window of interest and the remaining sums run over all
RPs in an event
\begin{eqnarray}
\mathcal{M}'_{abcd\cdots} &\equiv&
\sum_{i=1}^{m_p}\,\sump_{j,k,l,\ldots=1}^{M}\!\!\!\!\!w_i^a w_j^b
w_k^c w_l^d \cdots \,. 
\label{Mprimeabcd}
\end{eqnarray}

Using the definitions presented above 
the {\it weighted} single-event 2- and 4-particle
correlations are given by: 
\begin{eqnarray}
\mean{2}&\equiv&\!\!\frac{1}{\mathcal{M}_{11}}\,
\sump_{i,j=1}^{M}w_iw_j\,e^{in(\phi_i-\phi_j)}\,,
\label{2pCorrelationSingleEventGeneralParticleWeights}\\
\left<4\right>&\equiv&\!\!\frac{1}{\mathcal{M}_{1111}}\,
\sump_{i,j,k,l=1}^{M} \!\!\!\!\!w_iw_jw_kw_l\,
e^{in(\phi_i\!+\!\phi_j\!-\!\phi_k\!-\!\phi_l)}.
\label{4pCorrelationSingleEventGeneralParticleWeights}
\end{eqnarray}
The event weights (\ref{W<2>}) and (\ref{W<4>}) now read
\begin{eqnarray}
W_{\mean{2}}&\equiv&\mathcal{M}_{11}\,,
\label{W<2>generalPW}\\
W_{\left<4\right>}&\equiv&\mathcal{M}_{1111}
\label{W<4>generalPW}\,.
\end{eqnarray}
Analogously, the reduced single-event multi-particle correlations
now read:
\begin{eqnarray}
\left<2'\right>&\!\!\equiv&\!\!\!\frac{1}
{\mathcal{M}'_{01}}\,\sum_{i=1}^{m_{p}}
\sump_{j=1}^{M}w_j\,e^{in(\psi_i-\phi_j)},
\label{2pReducedCorrelationSingleEventGeneralPW}\\
\left<4'\right>&\!\!\equiv&\!\!\!\frac{1}{\mathcal{M}'_{0111}}\!
\sum_{i=1}^{m_p}\!
\sump_{j,k,l=1}^{M}\!\!\!\!w_jw_kw_l\,
e^{in(\psi_i\!+\!\phi_j\!-\!\phi_k\!-\!\phi_l)},
\label{4pReducedCorrelationSingleEventGeneralPW}
\end{eqnarray}
where the event weights (\ref{W<2'>}) and (\ref{W<4'>}) are now:
\begin{eqnarray}
w_{\left<2'\right>}&\equiv&\mathcal{M}'_{01}\,,
\nonumber \\
w_{\left<4'\right>}&\equiv&\mathcal{M}'_{0111}\,.
\end{eqnarray}
The weighted average 2-particle correlations are given
by the following equations:   
\begin{eqnarray}
\mean{2} &=& \frac{\left|Q_{n,1}\right|^2-S_{1,2}}{S_{2,1}-S_{1,2}}\,,
\nonumber\\
\left<\mean{2}\right> &=& \frac{\sum_{i=1}^N 
(\mathcal{M}_{11})_i \mean{2}_i}{\sum_{i=1}^N (\mathcal{M}_{11})_i}\,,
\nonumber\\
\mathcal{M}_{11} &\equiv& 
\sump_{i,j=1}^{M} w_i w_j
\nonumber\\
                 &=& S_{2,1}-S_{1,2}\,,
\label{2pWithWeights}
\end{eqnarray}
and the weighted average 4-particle correlations are given by: 
\begin{eqnarray}
\left<4\right>&=&\bigg[\left|Q_{n,1}\right|^4+
\left|Q_{2n,2}\right|^2-2\cdot
\mathfrak{Re}\left[Q_{2n,2}Q_{n,1}^*Q_{n,1}^*\right]
\nonumber\\
&+&8\cdot\mathfrak{Re}\left[Q_{n,3}Q_{n,1}^*\right]-
4\cdot S_{1,2}\left|Q_{n,1}\right|^2
\nonumber\\
&-& 6\cdot S_{1,4} - 2\cdot S_{2,2}\bigg] 
/\mathcal{M}_{1111}\,,
\nonumber\\
\mathcal{M}_{1111} &\equiv& \sump_{i,j,k,l=1}^{M} w_i w_j w_k w_l 
\nonumber\\
&=& S_{4,1} - 6\cdot S_{1,2}S_{2,1} + 8\cdot S_{1,3}S_{1,1}+3\cdot
S_{2,2}
\nonumber\\
&-& 6\cdot S_{1,4} \,,
\nonumber\\
\left<\left<4\right>\right>&=&\frac{\sum_{i=1}^N
  (\mathcal{M}_{1111})_i \left<4\right>_i}
{\sum_{i=1}^N (\mathcal{M}_{1111})_i}\,,
\label{4pWithWeights}
\end{eqnarray}
where the weighted $Q$-vector, $Q_{n,k}$, was defined in
Eq. (\ref{weightedQvector}) and $S_{p,k}$ in Eq. (\ref{Spk}). 

Weighted reduced 2- and 4-particle azimuthal correlations 
are given by the following formulas: 
\begin{eqnarray}
\left<2'\right> &=& \frac{p_{n,0} Q_{n,1}^* - s_{1,1}}{m_p
  S_{1,1}-s_{1,1}}\,,
\nonumber\\
\left<\left<2'\right>\right> &=&  \frac{\sum_{i=1}^N 
(\mathcal{M}'_{01})_i \left<2'\right>_i}{\sum_{i=1}^N 
(\mathcal{M}'_{01})_i}\,,
\nonumber\\
\mathcal{M}'_{01} &\equiv& \sum_{i=1}^{m_p}
  \sump_{i,j=1}^{M} w_j=m_p S_{1,1}-s_{1,1}\,,
\label{2pWithWeightsDiffFlow}
\end{eqnarray}
and, 
\begin{eqnarray}
\left<4'\right> &=& \bigg[ p_{n,0}Q_{n,1}Q_{n,1}^{*}Q_{n,1}^{*}
\nonumber\\
&-&q_{2n,1}Q_{n,1}^{*}Q_{n,1}^{*}-p_{n,0}Q_{n,1}Q_{2n,2}^{*}
\nonumber\\
&-&2\cdot S_{1,2} p_{n,0}Q_{n,1}^{*}-2\cdot s_{1,1}
\left|Q_{n,1}\right|^2
\nonumber\\
&+&7\cdot q_{n,2}Q_{n,1}^{*} - Q_{n,1}q_{n,2}^{*}
\nonumber\\
&+&q_{2n,1}Q_{2n,2}^{*}+2\cdot p_{n,0}Q_{n,3}^{*}\nonumber\\
&+&2\cdot s_{1,1}S_{1,2}-6\cdot s_{1,3}\bigg] 
/ \mathcal{M}'_{0111}\,, \nonumber\\
\left<\left<4'\right>\right>&\!\!\!\!=&\!\!\!\!\frac{\sum_{i=1}^N 
(\mathcal{M}'_{0111})_i \left<4'\right>_i}
{\sum_{i=1}^N (\mathcal{M}'_{0111})_i}\,,
\nonumber\\
\mathcal{M}'_{0111}&\!\!\!\!\equiv&\!\!\!\! 
\sum_{i=1}^{m_p} \sump_{j,k,l=1}^{M} w_j w_k w_l 
\nonumber\\
&\!\!\!\!=&\!\!\!\!m_p\left[S_{3,1}-3\cdot S_{1,1}S_{1,2}
+2\cdot S_{1,3}\right] 
\nonumber\\
&\!\!\!\!-&\!\!\!\! 3\!\cdot\!\left[s_{1,1}(S_{2,1}\!-\!S_{1,2})\!+\!2\!\cdot
\!(s_{1,3}\!-\!s_{1,2}S_{1,1})\right]\,.
\label{4pWithWeightsDiffFlow}
\end{eqnarray}
We note that to evaluate all quantities appearing on the right hand
sides in analytic expressions
(\ref{2pWithWeights}--\ref{4pWithWeightsDiffFlow}) only a single loop
over data is required.

\section{Non-uniform acceptance
\label{aNUA} 
}

Building cumulants from multi-particle correlations we have so far
omitted terms which vanish for the detectors with uniform acceptance. 
For a more general case they have to be 
kept~\cite{Borghini:2000sa,Borghini:2001vi,Selyuzhenkov:2007zi,K:1962}. 
The more general $2^{\mathrm{nd}}$ order cumulant now reads: 
\be
c_{n}\{2\} &=& \dmean{2} - \nonumber \\ 
& & \left[ \left<\left<\cos n\phi_1\right>\right>^2 
+ \left<\left<\sin n\phi_1\right>\right>^2 \right]\,.
\label{gen2ndQC}
\ee
The correction terms can be expressed in terms of the real 
and imaginary parts of the $Q$-vector
(\ref{Qvector}): 
\begin{eqnarray}
\left<\left<\cos n\phi_1\right>\right> &=& \frac{\sum_{i=1}^N
  \left(\mathfrak{Re} \left[Q_n\right]\right)_i}{\sum_{i=1}^N M_i}\,,
\label{cosPhi}\\
\left<\left<\sin n\phi_1\right>\right> &=& \frac{\sum_{i=1}^N
  \left(\mathfrak{Im} \left[Q_n\right]\right)_i}{\sum_{i=1}^N M_i}\,.
\label{sinPhi}
\end{eqnarray}
When particle weights are used the
average 2-particle correlation $\left<\mean{2}\right>$ is
determined from Eqs. (\ref{2pWithWeights}), while Eqs. (\ref{cosPhi})
and (\ref{sinPhi}) generalize into: 
\begin{eqnarray}
\left<\left<\cos n\phi_1\right>\right> &=& \frac{\sum_{i=1}^N
  \left(\mathfrak{Re} \left[Q_{n,1}\right]\right)_i}{\sum_{i=1}^N
  (S_{1,1})_i}\,,\label{cosPhiWithPW}\\ 
\left<\left<\sin n\phi_1\right>\right> &=& \frac{\sum_{i=1}^N
  \left(\mathfrak{Im} \left[Q_{n,1}\right]\right)_i}{\sum_{i=1}^N
  (S_{1,1})_i}\,,\label{sinPhiWithPW} 
\end{eqnarray}
where $Q_{n,1}$ can be determined from the definition of the weighted
$Q$-vector (\ref{weightedQvector}) and $S_{1,1}$ from definition
(\ref{Spk}). 
 
The generalized $4^{\mathrm{th}}$ order cumulant reads: 
\begin{eqnarray}
c_{n}\{4\} &=& \left<\left<4\right>\right> - 
2\cdot\left<\mean{2}\right>^2 -
\nonumber\\
&-& 4\cdot\left<\left<\cos
  n\phi_1\right>\right>\left<\left<\cos
  n(\phi_1-\phi_2-\phi_3)\right>\right>     
\nonumber\\
&+& 4\cdot\left<\left<\sin
  n\phi_1\right>\right>\left<\left<\sin
  n(\phi_1-\phi_2-\phi_3)\right>\right>      
\nonumber\\
&-&  \left<\left<\cos
  n(\phi_1+\phi_2)\right>\right>^2 - \left<\left<\sin
  n(\phi_1+\phi_2)\right>\right>^2
\nonumber\\
&+& 4\cdot
\left<\left<\cos n(\phi_1+\phi_2)\right>\right>
\nonumber\\
&\times& \left[\left<\left<\cos
    n\phi_1\right>\right>^2 - \left<\left<\sin
    n\phi_1\right>\right>^2\right]
\nonumber\\ 
&+& 8\cdot \left<\left<\sin n(\phi_1+\phi_2)\right>\right>
\left<\left<\sin n\phi_1\right>\right>
\left<\left<\cos n\phi_1\right>\right>
\nonumber\\
&+& 8 \cdot\left<\left<\cos n(\phi_1-\phi_2)\right>\right>
\nonumber\\
&\times& \left[\left<\left<\cos
    n\phi_1\right>\right>^2 + \left<\left<\sin
    n\phi_1\right>\right>^2\right]
\nonumber\\ 
& -& 6 \cdot \left[\left<\left<\cos n\phi_1\right>\right>^2 
+ \left<\left<\sin n\phi_1\right>\right>^2\right]^2\,.
\label{gen4thQC}
\end{eqnarray}
The terms starting from the second line in Eq. (\ref{gen4thQC})
counter balance 
the bias coming from non-uniform acceptance so that $c_{n}\{4\}$ is
unbiased. 
These terms can be expressed in terms of $Q$-vectors:
\begin{eqnarray}
\left<\left<\cos n(\phi_1\!+\!\phi_2)\right>\right> 
\!\!&=&\!\! \frac{\sum_{i=1}^N 
\left(\mathfrak{Re} \left[Q_nQ_n\! -\! Q_{2n}\right]\right)_i}
{\sum_{i=1}^N M_i(M_i\! -\! 1)},
\label{2pAnizCos}\\
\left<\left<\sin n(\phi_1\!+\!\phi_2)\right>\right> 
\!\!&=&\!\! \frac{\sum_{i=1}^N 
\left(\mathfrak{Im} \left[Q_nQ_n\! -\! Q_{2n}\right]\right)_i}
{\sum_{i=1}^N M_i(M_i\! -\! 1)},\label{2pAnizSin}
\end{eqnarray}
\begin{eqnarray}
\left<\left<\cos n(\phi_1\!-\!\phi_2\!-\!\phi_3)\right>\right> 
&\!\!=&\!\! \bigg\{\sum_{i=1}^N 
\left(\mathfrak{Re} \left[Q_nQ_n^*Q_n^* - Q_nQ_{2n}^*\right]\right.
\nonumber\\
&\hspace{-3.5cm}&\hspace{-3.5cm}\left. -2(M\!-\!1)
\mathfrak{Re} \left[Q_n^*\right]\right)_i\bigg\}
/ \sum_{i=1}^N M_i(M_i\! -\! 1)(M_i\!-\!2)\,,
\label{3pCos}\\
\left<\left<\sin n(\phi_1\!-\!\phi_2\!-\!\phi_3)\right>\right> 
&\!\!=&\!\! \bigg\{\sum_{i=1}^N 
\left(\mathfrak{Im} \left[Q_nQ_n^*Q_n^* - Q_nQ_{2n}^*\right]\right.
\nonumber\\
&\hspace{-3.5cm}&\hspace{-3.5cm}\left. -2(M\!-\!1)
\mathfrak{Im} \left[Q_n^*\right]\right)_i\bigg\}
/ \sum_{i=1}^N M_i(M_i\! -\! 1)(M_i\!-\!2)
\label{3pSin}\,.
\end{eqnarray}
When particle weights are used the average 2-particle 
correlation $\left<\mean{2}\right>$ is determined 
from Eqs. (\ref{2pWithWeights}), the average 4-particle 
correlation $\left<\left<4\right>\right>$ is determined from 
Eqs. (\ref{4pWithWeights}), the Eqs. (\ref{2pAnizCos}) 
and (\ref{2pAnizSin}) generalize into:
\begin{eqnarray}
\left<\left<\cos n(\phi_1\!+\!\phi_2)\right>\right> 
&=& \frac{\sum_{i=1}^N \left(\mathfrak{Re} \left[Q_{n,1}Q_{n,1} 
- Q_{2n,2}\right]\right)_i}{\sum_{i=1}^N (\mathcal{M}_{11})_i}\,,
\nonumber\\
\left<\left<\sin n(\phi_1\!+\!\phi_2)\right>\right> 
&=& \frac{\sum_{i=1}^N \left(\mathfrak{Im} \left[Q_{n,1}Q_{n,1} 
- Q_{2n,2}\right]\right)_i}{\sum_{i=1}^N (\mathcal{M}_{11})_i}\,,
\nonumber\\
\mathcal{M}_{11} &\equiv& \sump_{i,j=1}^{M}w_i w_j
                 = S_{2,1}-S_{1,2}\,,
\end{eqnarray}
and the Eqs. (\ref{3pCos}) and (\ref{3pSin}) generalize into
\begin{eqnarray}
\left<\left<\cos n(\phi_1\!-\!\phi_2\!-\!\phi_3)\right>\right> 
&\!\!=&\!\!
\bigg\{\sum_{i=1}^N \left(\mathfrak{Re} 
\left[Q_{n,1}Q_{n,1}^*Q_{n,1}^*\right.\right.
\nonumber\\
&\hspace{-3.64cm}&\hspace{-3.64cm}\left.\left. 
\!- Q_{n,1}Q_{2n,2}^*\!-\!2\!\cdot\! S_{1,2}Q_{n,1}^*\!+\!2
\!\cdot\! Q_{n,3}^*\right]\right)_i\bigg\}/\sum_{i=1}^N 
(\mathcal{M}_{111})_i\,,
\nonumber\\
\left<\left<\sin n(\phi_1\!-\!\phi_2\!-\!\phi_3)\right>\right> 
&\!\!=&\!\! \bigg\{\sum_{i=1}^N \left(\mathfrak{Im} 
\left[Q_{n,1}Q_{n,1}^*Q_{n,1}^* \right.\right.
\nonumber\\
&\hspace{-3.64cm}&\hspace{-3.64cm}\left.\left.\!- Q_{n,1}Q_{2n,2}^*\!
-\!2\!\cdot\! S_{1,2}Q_{n,1}^*\!+\!2\!
\cdot\! Q_{n,3}^*\right]\right)_i\bigg\}/\sum_{i=1}^N 
(\mathcal{M}_{111})_i\,,\nonumber\\
\mathcal{M}_{111} \!\equiv\!\! \!\!\!\sump_{i,j,k=1}^{M}\!\!\!\!\!w_i w_j w_k
&\!\!\!=&\!\!\! S_{3,1}\!-\!3\!\cdot\! S_{1,2}S_{1,1}\!+\!2\cdot\! S_{1,3}\,.
\end{eqnarray}

The generalized $2^{\mathrm{nd}}$ order differential cumulant reads
\begin{eqnarray}
& d_{n}\{2\} = \left<\left<2'\right>\right> - \nonumber \\
&\!\! \left<\left<\cos n\psi_1\right>\right>
\left<\left<\cos n\phi_2\right>\right>
\! - \!\dmean{\sin n\psi_1} \left<\left<\sin n\phi_2\right>\right>\,.
\label{gen2ndQC'}
\end{eqnarray}
Expressions for $\left<\left<\cos n\phi_1\right>\right>$ and
$\left<\left<\sin n\phi_1\right>\right>$ were already given in
Eqs. (\ref{cosPhi}) and (\ref{sinPhi}), respectively (when particle
weights are being used in Eqs. (\ref{cosPhiWithPW}) and
(\ref{sinPhiWithPW}), respectively). 
Similarly:
\begin{eqnarray}
\left<\left<\cos n\psi_1\right>\right> &=& \frac{\sum_{i=1}^N
  \left(\mathfrak{Re} \left[p_n\right]\right)_i}{\sum_{i=1}^N
  (m_p)_i}\,,
\label{cosPsi}\\
\left<\left<\sin n\psi_1\right>\right> &=& \frac{\sum_{i=1}^N
  \left(\mathfrak{Im} \left[p_n\right]\right)_i}{\sum_{i=1}^N
  (m_p)_i}\,,
\label{sinPsi}
\end{eqnarray}
where $p_n$ and $m_p$ were defined in Section~ \ref{sec:diffFlow}. The
Eqs. (\ref{cosPsi}) and (\ref{sinPsi}) remain unchanged when particle
weights are being used. 

The generalized $4^{\mathrm{th}}$ order differential cumulant reads:
\begin{eqnarray}
d_{n}\{4\} &=& \dmean{4'} - 2\cdot \left<\left<2'\right>\right>
\left<\mean{2}\right> 
\label{gen4thQC'}\\
&-& \left<\left<\cos n\psi_1\right>\right>
\left<\left<\cos n(\phi_1\!-\!\phi_2\!-\!\phi_3)\right>\right>
\nonumber\\
&+& \left<\left<\sin n\psi_1\right>\right>
\left<\left<\sin n(\phi_1\!-\!\phi_2\!-\!\phi_3)\right>\right>
\nonumber\\
&-& \left<\left<\cos n\phi_1\right>\right>
\left<\left<\cos n(\psi_1\!-\!\phi_2\!-\!\phi_3)\right>\right>
\nonumber\\
&+& \left<\left<\sin n\phi_1\right>\right>
\left<\left<\sin n(\psi_1\!-\!\phi_2\!-\!\phi_3)\right>\right>
\nonumber\\
&-& 2\cdot\left<\left<\cos n\phi_1\right>\right>
\left<\left<\cos n(\psi_1\!+\!\phi_2\!-\!\phi_3)\right>\right>
\nonumber\\
&-& 2\cdot\left<\left<\sin n\phi_1\right>\right>
\left<\left<\sin n(\psi_1\!+\!\phi_2\!-\!\phi_3)\right>\right>
\nonumber\\
&-&\left<\left<\cos n(\psi_1\!+\!\phi_2)\right>\right>
\left<\left<\cos n(\phi_1\!+\!\phi_2)\right>\right>
\nonumber\\
&-&\left<\left<\sin n(\psi_1\!+\!\phi_2)\right>\right>
\left<\left<\sin n(\phi_1\!+\!\phi_2)\right>\right>
\nonumber\\
&+& 2\cdot\left<\left<\cos n(\phi_1+\phi_2)\right>\right>
\nonumber\\
&\times& \left[\left<\left<\cos n\psi_1\right>\right>
\left<\left<\cos n\phi_1\right>\right>
\! -\! \left<\left<\sin n\psi_1\right>\right>
\left<\left<\sin n\phi_1\right>\right> \right] 
\nonumber\\
&+&2\cdot\left<\left<\sin n(\phi_1\!+\!\phi_2)\right>\right>
\nonumber\\
&\times&\left[\left<\left<\cos n\psi_1\right>\right>
\left<\left<\sin n\phi_1\right>\right>
\! +\! \left<\left<\sin n\psi_1\right>\right>
\left<\left<\cos n\phi_1\right>\right> \right] 
\nonumber\\
&+& 4\cdot\left<\left<\cos n(\phi_1\!-\!\phi_2)\right>\right>
\nonumber\\
&\times& \left[\left<\left<\cos n\psi_1\right>\right>
\left<\left<\cos n\phi_1\right>\right>
\! +\! \left<\left<\sin n\psi_1\right>\right>
\left<\left<\sin n\phi_1\right>\right> \right] 
\nonumber\\
&+& 2\cdot\left<\left<\cos n(\psi_1\!+\!\phi_2)\right>\right>
\nonumber\\
&\times&\left[\left<\left<\cos n\phi_1\right>\right>^2 
\!-\! \left<\left<\sin n\phi_1\right>\right>^2 \right] 
\nonumber\\
&+& 4\cdot\left<\left<\sin n(\psi_1\!+\!\phi_2)\right>\right> 
\left<\left<\cos n\phi_1\right>\right> 
\left<\left<\sin n\phi_1\right>\right>  \nonumber\\
&+& 4\cdot\left<\left<\cos n(\psi_1\!-\!\phi_2)\right>\right>
\left[\left<\left<\cos n\phi_1\right>\right>^2 
\!+\! \left<\left<\sin n\phi_1\right>\right>^2 \right] 
\nonumber\\
&-& 6\cdot\left[\left<\left<\cos n\phi_1\right>\right>^2
\! -\! \left<\left<\sin n\phi_1\right>\right>^2 \right]
\nonumber \\
&\times&\left[\left<\left<\cos n\psi_1\right>\right>
\left<\left<\cos n\phi_1\right>\right>
\! -\! \left<\left<\sin n\psi_1\right>\right>
\left<\left<\sin n\phi_1\right>\right> \right] 
\nonumber\\
&-& 12\cdot\left<\left<\cos n\phi_1\right>\right> 
\left<\left<\sin n\phi_1\right>\right>
\nonumber\\
&\times& \left[\left<\left<\sin n\psi_1\right>\right>
\left<\left<\cos n\phi_1\right>\right> 
\!+\! \left<\left<\cos n\psi_1\right>\right>
\left<\left<\sin n\phi_1\right>\right> \right] 
\nonumber\,.
\end{eqnarray}
The terms starting from the second line in Eq. (\ref{gen4thQC'}) counter
balance the bias coming from non-uniform acceptance. 
Some of the new terms appearing in this expression can be expressed 
again in products of flow vectors:
\begin{eqnarray}
\left<\left<\cos n(\psi_1\!+\!\phi_2)\right>\right> 
&\!\!=&\!\! \frac{\sum_{i=1}^N\!\! 
\left(\mathfrak{Re} \left[p_nQ_n \!- \!q_{2n}\right]\right)_i}
{\sum_{i=1}^N \!\!\left(m_p M\!-\!m_q\right)_i}\,,
\nonumber\\
\left<\left<\sin n(\psi_1\!+\!\phi_2)\right>\right> 
&\!\!=&\!\! \frac{\sum_{i=1}^N \!\!\left(\mathfrak{Im} 
\left[p_nQ_n\! - \!q_{2n}\right]\right)_i}{\sum_{i=1}^N 
\!\!\left(m_p M\!-\!m_q\right)_i}\,,
\label{2pAnizDF}
\end{eqnarray}
\begin{eqnarray}
\left<\left<\cos n(\psi_1\!+\!\phi_2-\!\phi_3)\right>\right> 
&=& \bigg\{\sum_{i=1}^N \big(\mathfrak{Re} \left[p_n\!\left
(\left|Q_n\right|^2\!-\!M\right)\right]
\nonumber\\
&\hspace{-3.84cm}&\hspace{-3.84cm} -\mathfrak{Re} 
\left[q_{2n}Q_n^*\!+\!m_qQ_n\!-\!2q_n\right]\big)_i\bigg\}
/ \sum_{i=1}^N \left[(m_pM\!-\!2m_q)(M\!-\!1)\right]_i\,,
\nonumber\\
\left<\left<\sin n(\psi_1\!+\!\phi_2-\!\phi_3)\right>\right> 
&=& \bigg\{\sum_{i=1}^N \big(\mathfrak{Im} \left[p_n\!
\left(\left|Q_n\right|^2\!-\!M\right)\right]
\nonumber\\
&\hspace{-3.84cm}&\hspace{-3.84cm} -\mathfrak{Im} 
\left[q_{2n}Q_n^*\!+\!m_qQ_n\!-\!2q_n\right]\big)_i\bigg\}
/ \sum_{i=1}^N \left[(m_pM\!-\!2m_q)(M\!-\!1)\right]_i\,,
\nonumber\\
\label{3pAnizDFa}
\end{eqnarray}
\begin{eqnarray}
\left<\left<\cos n(\psi_1\!-\!\phi_2-\!\phi_3)\right>\right> 
&=& \bigg\{\sum_{i=1}^N \big(\mathfrak{Re} 
\left[p_nQ_n^*Q_n^*\!-\!p_nQ_{2n}^*\right]
\nonumber\\
&\hspace{-3.84cm}&\hspace{-3.84cm} -\mathfrak{Re} 
\left[2m_qQ_n^*\!-\!2q_n^*\right]\big)_i\bigg\}
/ \sum_{i=1}^N \left[(m_pM-2m_q)(M-1) \right]_i\,,
\nonumber\\
\left<\left<\sin n(\psi_1\!-\!\phi_2-\!\phi_3)\right>\right> 
&=& \bigg\{\sum_{i=1}^N \big(\mathfrak{Im} 
\left[p_nQ_n^*Q_n^*-p_nQ_{2n}^*\right]
\nonumber\\
&\hspace{-3.84cm}&\hspace{-3.84cm} -\mathfrak{Im} \left[
2m_qQ_n^*\!-\!2q_n^*\right]\big)_i
\bigg\}/\sum_{i=1}^N \left[(m_pM-2m_q)(M-1)\right]_i\,.
\nonumber\\
\label{3pAnizDFb}
\end{eqnarray}
When particle weights are used Eqs. (\ref{2pAnizDF})
generalize into: 
\begin{eqnarray}
\left<\left<\cos n(\psi_1\!+\!\phi_2)\right>\right> 
&=& \frac{\sum_{i=1}^N \left(\mathfrak{Re} \left[p_nQ_{n,k} 
- q_{2n,k}\right]\right)_i}{\sum_{i=1}^N 
\left(m_p S_{1,1}-s_{1,1}\right)_i}\,,
\nonumber\\
\left<\left<\sin n(\psi_1\!+\!\phi_2)\right>\right> 
&=& \frac{\sum_{i=1}^N \left(\mathfrak{Im} \left[p_nQ_{n,k} -
    q_{2n,k}\right]\right)_i}{\sum_{i=1}^N \left(m_p
  S_{1,1}-s_{1,1}\right)_i}\,,
\nonumber\\
\end{eqnarray}
Eqs. (\ref{3pAnizDFa}) generalize into:
\begin{eqnarray}
\left<\left<\cos n(\psi_1\!+\!\phi_2-\!\phi_3)\right>\right> &\!\!\!=&\!\!\! 
\bigg\{\sum_{i=1}^N \big(\mathfrak{Re}
\left[p_n\left(\left|Q_{n,1}\right|^2\!-\!S_{1,2}\right)\right]
\nonumber\\
&\hspace{-3.84cm}&\hspace{-3.84cm} -\mathfrak{Re}
\left[q_{2n,1}Q_{n,1}^*+s_{1,1}Q_{n,1}-2q_{n,2}\right]\big)_i\bigg\}/
\nonumber\\
&\hspace{-3.84cm}&\hspace{-3.84cm}\bigg\{\sum_{i=1}^N
\left(m_p(S_{2,1}-S_{1,2})-2\cdot(s_{1,1} S_{1,1} -
s_{1,2})\right)_i\bigg\}\,,
\nonumber\\
\left<\left<\sin n(\psi_1\!+\!\phi_2-\!\phi_3)\right>\right>
&\!\!\!=&\!\!\!\bigg\{\sum_{i=1}^N \big(\mathfrak{Im}
\left[p_n\left(\left|Q_{n,1}\right|^2\!-\!S_{1,2}\right)\right]
\nonumber\\
&\hspace{-3.84cm}&\hspace{-3.84cm} -\mathfrak{Im}
\left[q_{2n,1}Q_{n,1}^*\!+\!s_{1,1}Q_{n,1}\!-\!2q_{n,2}\right]\big)_i\bigg\}/
\nonumber\\
&\hspace{-3.84cm}&\hspace{-3.84cm}\bigg\{\sum_{i=1}^N 
\left(m_p(S_{2,1}-S_{1,2})-2\cdot(s_{1,1} S_{1,1} -
s_{1,2})\right)_i\bigg\}\,, 
\end{eqnarray}
and finally, Eqs. (\ref{3pAnizDFb}) generalize into:
\begin{eqnarray}
\left<\left<\cos n(\psi_1\!-\!\phi_2-\!\phi_3)\right>\right> &\!\!\!=&\!\!\! 
\bigg\{\sum_{i=1}^N \big(\mathfrak{Re} 
\left[p_n\left(Q_{n,1}^*Q_{n,1}^*\!-\!Q_{2n,2}^*\right)\right]
\nonumber\\
&\hspace{-3.44cm}&\hspace{-3.44cm} -2\cdot\mathfrak{Re}\left[
  s_{1,1}Q_{n,1}^*-q_{n,2}^*\right]\big)_i\bigg\}/
\nonumber\\
&\hspace{-3.44cm}&\hspace{-3.44cm}\bigg\{\sum_{i=1}^N
\left(m_p(S_{2,1}-S_{1,2})-2\cdot(s_{1,1} S_{1,1} -
s_{1,2})\right)_i\bigg\}\,,
\nonumber\\
\left<\left<\sin n(\psi_1\!-\!\phi_2-\!\phi_3)\right>\right>
&\!\!\!=&\!\!\!\bigg\{\sum_{i=1}^N \big(\mathfrak{Im}
\left[p_n\left(Q_{n,1}^*Q_{n,1}^*\!-\!Q_{2n,2}^*\right)\right]
\nonumber\\
&\hspace{-3.44cm}&\hspace{-3.44cm} -2\cdot\mathfrak{Im}\left[
  s_{1,1}Q_{n,1}^*-q_{n,2}^*\right]\big)_i\bigg\}/
\nonumber\\
&\hspace{-3.44cm}&\hspace{-3.44cm}\bigg\{\sum_{i=1}^N
\left(m_p(S_{2,1}-S_{1,2})-2\cdot(s_{1,1} S_{1,1} -
s_{1,2})\right)_i\bigg\}\,. 
\end{eqnarray}


\end{document}